\documentclass[journal,twocolumn,twoside]{IEEEtran}

\usepackage{amsfonts}
\usepackage{amsmath}
\usepackage{amssymb}
\usepackage{xcolor}
\usepackage{bbm}
\usepackage{cite}
\usepackage{graphics} 
\usepackage{graphicx}
\usepackage{xcolor}
\usepackage{subfigure}

\usepackage{stfloats}

\usepackage{lipsum}

\newtheorem{theorem}{Theorem}

\newtheorem{corollary}[theorem]{Corollary}

\newtheorem{definition}[theorem]{Definition}

\newtheorem{lemma}[theorem]{Lemma}

\newtheorem{proposition}[theorem]{Proposition}
\newtheorem{remark}[theorem]{Remark}

\newenvironment{proof}[1][Proof]{\textbf{#1.} }{\ \rule{0.5em}{0.5em}}

\def\norm #1{\left\|#1\right\|}

\def\frobn #1{\left\|#1\right\|_{\text{F}}}

\def\abs #1{\left|#1\right|}

\def\bC{\mathbb{C}}

\def\bR{\mathbb{R}}
\def\bE{\mathbb{E}}

\def\m #1{\boldsymbol{#1}}

\def\cP{\mathcal{P}}

\def\cR{\mathcal{R}}

\def\cT{\mathcal{T}}
\def\cU{\mathcal{U}}

\def\bee{\begin{equation}}
\def\ene{\end{equation}}

\def\beq{\begin{eqnarray}}
\def\enq{\end{eqnarray}}
\def\lentwo{\setlength\arraycolsep{2pt}}

\def\equ #1{\begin{equation}#1\end{equation}}
\def\equa #1{\begin{eqnarray}#1\end{eqnarray}}
\def\sbra#1{\left(#1\right)}
\def\mbra #1{\left[#1\right]}
\def\lbra #1{\left\{#1\right\}}
\def\diag #1{\text{diag}#1}

\def\rank#1{\text{rank}#1}

\begin{document}

\title{Nonasymptotic Performance Analysis of ESPRIT and Spatial-Smoothing ESPRIT}
\author{Zai Yang\thanks{IEEE Transactions on Information Theory, to appear

The author is with School of Mathematics and Statistics, Xi'an Jiaotong University, Xi'an 710049, China (e-mail: yangzai@xjtu.edu.cn).}}
\maketitle

\begin{abstract}
This paper is concerned with the problem of frequency estimation from multiple-snapshot data. It is well-known that ESPRIT (and spatial-smoothing ESPRIT in presence of coherent sources or given limited snapshots) can locate the true frequencies if either the number of snapshots or the signal-to-noise ratio (SNR) approaches infinity. In this paper, we analyze the nonasymptotic performance of ESPRIT and spatial-smoothing ESPRIT with finitely many snapshots and finite SNR. We show that the absolute frequency estimation error of ESPRIT (or spatial-smoothing ESPRIT) is bounded from above by $C\frac{\max\lbra{\sigma, \sigma^2}}{\sqrt{L}}$ with overwhelming probability, where $\sigma^2$ denotes the Gaussian noise variance, $L$ is the number of snapshots and $C$ is a coefficient independent of $L$ and $\sigma^2$, if and only if the true frequencies can be localized by ESPRIT (or spatial-smoothing ESPRIT) without noise or with infinitely many snapshots. Our results are obtained by deriving new matrix perturbation bounds and generalizing the classical Schur product theorem, which may be of independent interest. Extensions to MUSIC and spatial-smoothing MUSIC are also made. Numerical results are provided corroborating our analysis.
\end{abstract}


\section{Introduction}
Frequency estimation from a single or multiple snapshots of the superposition of several sinusoidal waves is a fundamental problem in statistical signal processing and has broad applications in array, radar and sonar signal processing, wireless communications, structural health monitoring, etc \cite{stoica2005spectral}. The Nyquist-Shannon sampling theorem states that a sampling rate at least twice the highest frequency is required for lossless information recovery. In the practical scenario of finitely many noisy samples, algorithms have been constantly developed for accurate frequency estimation, which range from the classical periodogram/beamformer to subspace-based methods since 1980s and then to sparse and compressed sensing approaches in this century \cite{krim1996two,yang2018sparse}. Correspondingly, the computations of these methods shift from the Fast Fourier Transform (FFT) to eigenvalue decompositions and then to iterative optimization algorithms thanks to the continuous improvement of computing power. Extensive studies over the last four decades have witnessed excellent performance of subspace methods. This paper is devoted to a nonasymptotic analysis of subspace methods showing their stability and high resolution when the samples per snapshot are just sufficient to do so.



The multiple snapshots of data sequence acquired at a Nyquist sampling rate can be modeled as \cite{stoica2005spectral}:
\equ{\m{Y} = \m{A}\m{S} + \m{E}, \label{eq:model}}
where $\m{Y}$ is an $N\times L$ matrix of which every column corresponds to one snapshot, $\m{A}$ is an $N\times K$ Vandermonde matrix whose $\sbra{n,k}$ entry is $e^{i2\pi\sbra{n-1}f_k}$ with $i=\sqrt{-1}$, $\m{S}$ is a $K\times L$ matrix, and $\m{E}$ denotes noise. The objective is to estimate the set of distinct frequencies $\cT = \lbra{f_k\in[0,1)}_{k=1}^K$ given $\m{Y}$ and $K$. While the above frequency estimation problem arises in many applications, we will use the language of array signal processing for convenience in this paper. In array processing, we need to estimate the directions $\lbra{\theta_k\in\left[-90^\circ, 90^\circ\right)}$ of $K$ narrowband, far-field sources impinging on an $N$-element uniform linear array (ULA) from $L$-snapshot outputs $\m{Y}$ of the array, known as direction-of-arrival (DOA) estimation. The DOAs $\lbra{\theta_k}$ are connected to the frequencies $\lbra{f_k}$ by $f_k = \frac{d}{\lambda}\sin\theta_k \text{ mod }1$, where $d$ denotes the distance between adjacent antennas, $\lambda$ is the wavelength and the modulo operation is nothing special but keeps $f_k\in[0,1)$ (note that the matrix $\m{A}$ is invariant with the modulo operation). A typical assumption is that the distance $d$ is half a wavelength, i.e., $d = \frac{\lambda}{2}$, so that the mapping between $\theta_k$ and $f_k$ is one-to-one and $\theta_k$ is uniquely determined by $f_k$ with
\equ{\theta_k = \left\{\begin{array}{ll} \arcsin\sbra{2f_k}, & f_k \in \left[0,\frac{1}{2}\right),\\ \arcsin\sbra{2f_k-2}, & f_k \in \left[\frac{1}{2},1\right).\end{array} \right. \notag}
The Vandermonde matrix $\m{A}$ is referred to as the array manifold matrix. The $K\times L$ matrix $\m{S}$ consists of emitting signals of the $K$ sources at $L$ snapshots. Consider the $K$ sources as random processes. Then, we use the terminologies of independent, uncorrelated, correlated and coherent (fully correlated) sources without ambiguity. Note that coherent sources can be caused by multipath propagations of emitting sources that, as detailed later, bring challenges to subspace methods.

It is seen from \eqref{eq:model} that the sampled data $\m{Y}$ are highly nonlinear functions of the frequencies $\lbra{f_k}$ of interest. To overcome the nonlinearity and to circumvent nonconvex optimizations, subspace-based methods are proposed by observing that the frequencies can be uniquely identified from the range space of $\m{A}$ that corresponds to the eigen-subspace, associated with the greatest $K$ eigenvalues, of the data covariance matrix $\m{R}$ of each snapshot. The aforementioned subspace is usually referred to as the signal subspace and its orthogonal subspace is known as the noise subspace. In the practical scenario with finitely many snapshots, $\m{R}$ is replaced by its efficient estimate, the sample covariance matrix $\widehat{\m{R}}=\frac{1}{L}\m{Y}\m{Y}^H$. Based on the above principle, two prominent subspace methods are multiple signal classification (MUSIC) \cite{schmidt1981signal,schmidt1986multiple} (by using the noise subspace) and estimation of signal parameters via rotational invariant techniques (ESRPIT) \cite{roy1986esprit,paulraj1986subspace, roy1989esprit} (by using the signal subspace). A partial list of their variants includes root-MUSIC \cite{barabell1983improving}, unitary-ESPRIT \cite{haardt1995unitary}, ESPRIT with fourth-order statistics \cite{chiang1989esprit}, as well as their extensions to higher dimensions \cite{haardt19952d, li1992two,hua1993pencil,hua1992estimating}.

One prerequisite for the use of MUSIC and ESPRIT for frequency estimation is that the source signal matrix $\m{S}$ has full row rank so that the $K$-dimensional signal subspace can be retrieved from the data covariance matrix $\m{R}$, which however fails to hold in the presence of coherent sources or given a small number of snapshots. Spatial smoothing (SS), which was pioneered in \cite{evans1981high,evans1982application} and has been extensively studied since then (see, e.g., \cite{shan1985spatial,shan1987smoothed,williams1988improved, pillai1989forward,du1991improved, friedlander1992direction,li1992improved, rao1993weighted,thompson1996performance, tan1997estimating,wang19982, choi2002conditions,qi2005spatial,thakre2009single, thakre2010tensor,dai2011spatial,yang2019source}), is an effective preprocessing technique to restore the performance of subspace methods in this case by leveraging the effective array aperture and the rank of smoothed source signal matrix. Two typical SS techniques are forward-only SS (FOSS) and forward-backward SS (FBSS). The resulting algorithms when combined with MUSIC and ESPRIT is called SS-MUSIC and SS-ESPRIT that encompass the standard MUSIC and ESPRIT as special cases.

It is well-known that MUSIC, ESPRIT and their SS counterparts produce the true frequencies if the number of snapshots $L$ or the signal-to-noise ratio (SNR) approaches infinity (the latter case corresponds to the noiseless case) under mild conditions (see details in the main context). In fact, an FBSS-based subspace method does so if and only if the frequencies can be uniquely identified from the data \cite{bresler1986number}. This paper is concerned with the practical scenario with finitely many snapshots and finite SNR. Our main result is stated as follows:
\begin{theorem} (Informal) Under Gaussianity assumptions on the sources and noise, the absolute frequency estimation error of ESPRIT (or SS-ESPRIT) is upper bounded by $C\frac{\max\lbra{\sigma, \sigma^2}}{\sqrt{L}}$ with overwhelming probability, where $\sigma^2$ denotes the noise variance and $C$ is a problem-dependent coefficient that is independent of $\sigma^2$ and $L$, if and only if ESPRIT (or SS-ESPRIT) produces the true frequencies in the limiting case of $\sigma\rightarrow 0$ or $L\rightarrow \infty$. \label{thm:informal}
\end{theorem}

It is implied by Theorem \ref{thm:informal} that both ESPRIT and SS-ESPRIT can stably estimate the frequencies provided that $\frac{\max\lbra{\sigma, \sigma^2}}{\sqrt{L}}$ is smaller than a certain threshold. Therefore, there is no substantial performance gap between the practical scenario of finite $L$ and finite SNR and the limiting case of infinite $L$ or infinite SNR. Our results also generalize to MUSIC and SS-MUSIC.

Our technical analyses are based on a new matrix perturbation bound, the Hadamard product technique for SS in our recent work \cite{yang2019source}, and the results in \cite{liao2016music,fannjiang2016compressive,li2020super}. In particular, the signal subspace estimation step of ESPRIT and SS-ESPRIT is interpreted with principal component analysis \cite{abdi2010principal} and analyzed with matrix perturbation theory \cite{stewart1990matrix}. A new matrix perturbation bound is derived in order to show that the subspace estimation error is a decreasing function of the snapshot number $L$ as well as to deal with non-i.i.d.~noise arising due to SS. By applying the new bound and its technical proof, the subspace estimation errors of ESPRIT and SS-ESPRIT are quantified. In this process, the Hadamard product technique in \cite{yang2019source} plays an important role. To measure the frequency estimation error based on the subspace estimation error, we apply the results on single-snapshot MUSIC and ESPRIT derived in \cite{liao2016music,fannjiang2016compressive,li2020super}. Nontrivial lower bounds on the minimum eigenvalue of the Hadamard product of singular positive-semidefinite matrices are provided which generalizes the classical Schur product theorem \cite{schur1911bemerkungen} and helps to understand the derived error bounds when the per-snapshot sample size $N$ is small.

\subsection{Relations to Prior Art}
Extensive studies on the frequency estimation errors of MUSIC and ESPRIT have been carried out under the critical assumption that either $L$ or the SNR is sufficiently large so that the frequency solutions are close to their ground truth; however, there is no explicit quantification of how large they should be and therefore, it is not guaranteed that the derived results are applicable to any problem with finitely many snapshots and finite SNR. Such studies are known as asymptotic analysis; see, e.g., \cite{stoica1989music, rao1989performance1, rao1989performance, stoica1990music, stoica1991performance,ottersten1991performance, mathews1994performance,mathews1996performance, yuen1996asymptotic,steinwandt2017performance}. In contrast to this, our results are nonasymptotic in the sense that they are applicable to any $L$ or SNR provided that the other is greater than a given threshold.

The nonasymptotic analysis presented in this paper is closely related to \cite{aubel2016deterministic} and a line of papers by Fannjiang, Liao and Li \cite{liao2016music,fannjiang2016compressive, liao2015music,liao2014music, li2020super}. The nonasymptotic performance of MUSIC and ESPRIT in the single-snapshot case, which are FOSS-MUSIC and FOSS-ESPRIT in the language of this paper, is studied in \cite{liao2016music,fannjiang2016compressive,aubel2016deterministic, li2020super, liao2015music} when the sample size $N$ is greater than twice the number of frequencies $K$, where SS is adopted to compensate deficiency of the snapshots. Differently from the single-snapshot case, we resolve two important and challenging problems that arise in the multiple-snapshot case, to be specific, effects of the snapshot number $L$ and coherent sources. We provide an explicit error bound that is a decreasing function of $L$ and diminishes as $L$ increases to infinity. In the presence of coherent sources, we show similar results for FOSS-ESPRIT and FBSS-ESPRIT where SS is used to deal with coherent sources. When applied to the single-snapshot case, our result on FBSS-ESPRIT implies that stable frequency estimation can be obtained even when the sample size $N$ is smaller than twice the number of frequencies $K$. We note that MUSIC with multiple snapshots is studied in \cite{liao2014music} where, unlike this paper, nonuniform samples are considered and neither of the aforementioned problems are tackled.

Note that the main difference between the single-snapshot and multiple-snapshot MUSIC and ESPRIT is the way of estimating the signal subspace. Given the results on single-snapshot MUSIC and ESPRIT in \cite{liao2016music,fannjiang2016compressive, li2020super}, the remaining difficulty in nonasymptotic analysis of multiple-snapshot MUSIC and ESPRIT is to quantify the error of signal subspace estimation with respect to the number of snapshots $L$ and possibly in presence of coherent sources. This has been attempted in the preprint paper \cite{li2021stability} by assuming non-coherent sources; but unfortunately, its critical error bound in \cite[Theorem III.4]{li2021stability} is severely flawed (see detailed explanations in Appendix A). In contrast to this, we present a new matrix perturbation result in this paper to bound the signal subspace estimation error, which is also generalized to analyze FOSS-ESPRIT and FBSS-ESPRIT in presence of coherent sources.


Resolution is an important property of a frequency estimation method that measures to what extent closely located frequencies can be resolved. It is shown in \cite{candes2013towards,tang2013compressed,adcock2015generalized, yang2016exact,fernandez2016super} that a resolution of $\frac{4}{N}$ or $\frac{2.52}{N}$ can be achieved by recent atomic norm methods within the framework of infinite-dimensional or gridless compressed sensing regardless of coherent sources and the snapshot number $L$. The paper \cite{fannjiang2011exact} considers a conventional compressed sensing setup (by assuming that the point sources are located on discrete grid points) and uncorrelated sources (resulting from random illuminations) and shows that the resolution of $\ell_1$ minimization can be improved by increasing the snapshot number (corresponding to the number of random illuminations). MUSIC and ESPRIT are called high-resolution methods since when they were proposed because empirically they have a resolution higher than the Rayleigh length $\frac{1}{N}$ given sufficient snapshots. In this paper, we provide rigorous analysis for ESPRIT and SS-ESPRIT in the nonasymptotic setting and show that their resolution improves constantly (without a limit) as the number of snapshots or the SNR increases.

\subsection{Notation}

Notations used in this paper are as follows. The set of real and complex numbers are denoted $\bR$ and $\bC$ respectively. Boldface letters are reserved for vectors and matrices. The amplitude of scalar $a$ is denoted $\abs{a}$. The complex conjugate, transpose, complex transpose and pseudo-inverse of matrix $\m{A}$ are denoted $\overline{\m{A}}$, $\m{A}^T$, $\m{A}^H$ and $\m{A}^{\dag}$ respectively. The rank and spectral norm of matrix $\m{A}$ are denoted $\rank\sbra{\m{A}}$ and $\norm{\m{A}}$. The maximum, $j$th greatest, and the minimum eigenvalues (or singular values) of a matrix are denoted $\lambda_{\text{max}}\sbra{\cdot}$, $\lambda_j\sbra{\cdot}$ and $\lambda_{\text{min}}\sbra{\cdot}$ (or $\sigma_{\text{max}}\sbra{\cdot}$, $\sigma_j\sbra{\cdot}$ and $\sigma_{\text{min}}\sbra{\cdot}$). We write $\m{A}>\m{0}$ (or $\m{A}\geq\m{0}$) if $\m{A}$ is Hermitian and positive (semi)definite. The $j$th entry of vector $\m{x}$ is $x_j$, and the $(j,l)$ entry of matrix $\m{A}$ is $A_{jl}$. For vector $\m{x}$, $\diag\sbra{\m{x}}$ denotes a diagonal matrix with $\m{x}$ on the diagonal; for a square matrix $\m{A}$, $\diag\sbra{\m{A}}$ denotes a column vector composed of the diagonal entries of $\m{A}$. The Hadamard (or elementwise) product of matrices $\m{A},\m{B}$ is denoted $\m{A}\odot\m{B}$. The expectation of a random variable is denoted $\bE[\cdot]$.

\subsection{Organization}
The rest of the paper is organized as follows. We revisit ESPRIT and SS-ESPRIT in Section \ref{sec:esprit}. We present previous matrix perturbation bounds and derive a new result in Section \ref{sec:pertbound}. The signal subspace estimation errors of ESPRIT and SS-ESPRIT are measured by applying the new matrix perturbation bound in Section \ref{sec:subspaceest}. The error bounds for ESPRIT and SS-ESPRIT are provided in Section \ref{sec:bound} to analyze their stability and resolution in frequency estimation. Positive-definiteness and the minimum eigenvalue of a Hadamard product, which are important features of the derived error bounds, are studied in Section \ref{sec:hadamard}. Detailed proofs of several theorems and lemmas and extensions to MUSIC and SS-MUSIC are provided in Appendices.

\section{ESPRIT and SS-ESPRIT} \label{sec:esprit}

\subsection{Assumptions}
We will make one or more of the following assumptions throughout this paper:
\begin{itemize}
\item[\textbf{A1:}] The entries of $\m{E}$ are i.i.d.~complex Gaussian with zero mean and variance $\sigma^2$;
\item[\textbf{A2:}] The columns of $\m{S}$ are independently drawn from a complex Gaussian distribution with zero mean and covariance $\m{\Sigma}$;
\item[\textbf{A3:}] $\m{E}$ and $\m{S}$ are independent.
\end{itemize}
The Gaussianity assumptions on the sources and noise are commonly used in array signal processing. In the limiting case of $L\rightarrow \infty$ or $\sigma\rightarrow 0$, to be concerned in Theorems \ref{thm:ESPRIT} and \ref{thm:SS_ESPRIT}, they can be relaxed to any other distributions. They are made in our nonasymptotic analysis so that all constants involved are given explicitly, though they can be relaxed to subgaussian to show the same scaling behaviors with respect to the snapshot number $L$ and the noise level $\sigma$.

\subsection{ESPRIT}


In ESPRIT, the frequencies are estimated from the signal subspace that is computed from the sample covariance matrix given by
\equ{\widehat{\m{R}} = \frac{1}{L}\m{Y}\m{Y}^H.}
To understand how it works, let us consider the extreme case in which the number of snapshots $L$ approaches infinity and the sample covariance matrix equals the data covariance matrix almost surely that under assumptions A1--A3 is given by
\equ{\m{R} = \m{A}\m{\Sigma}\m{A}^H + \sigma^2\m{I}. \label{eq:R}}
If the source covariance matrix $\m{\Sigma}$ is positive definite, then $\m{A}\m{\Sigma}\m{A}^H$ is positive semidefinite and has exactly rank $K$. Let
\equ{\m{R} = \sum_{j=1}^N \lambda_j \m{u}_j \m{u}_j^H \label{eq:Reigdec}}
be the eigen-decomposition of $\m{R}$, where $\lbra{\lambda_j}$ are the eigenvalues sorted in descending order and thus satisfy
\equ{\lambda_1\geq \dots \geq \lambda_K > \lambda_{K+1} = \dots =\lambda_{N}=\sigma^2,}
and $\lbra{\m{u}_j}$ are the associated eigenvectors. Then, we divide the eigenvalues into two groups and write \eqref{eq:Reigdec} as
\equ{\m{R} = \m{U}\m{\Lambda}\m{U}^H + \sigma^2 \m{U}_{\perp}\m{U}_{\perp}^H, \label{eq:Reigdec2}}
where $\m{\Lambda} = \diag\lbra{\lambda_1,\dots,\lambda_K}$, $\m{U}$ is composed of the first $K$ eigenvectors, and $\m{U}_{\perp}$ is perpendicular to $\m{U}$ and consists of the other $N-K$ eigenvectors. It can easily be shown that $\m{U}$ and $\m{A}$ share the same range space that is referred to as the signal subspace. Its orthogonal subspace, the range space of $\m{U}_{\perp}$, is called the noise subspace. ESPRIT is an algorithm that identifies the frequencies from the signal subspace. In particular, let $\m{U}_1$ and $\m{U}_2$ be the submatrices of $\m{U}$ by removing its last and first row, respectively. It can be shown that the eigenvalues of the matrix $\m{U}_1^{\dag}\m{U}_2$ are exactly $z_k = e^{i2\pi f_k}$, $k=1,\dots,K$, from which $\lbra{f_k}$ are obtained.

In practice, we have only finitely many snapshots and the covariance estimate $\widehat{\m{R}}$. To estimate the frequencies, we compute the eigen-decomposition of $\widehat{\m{R}}$:
\equ{\widehat{\m{R}} = \sum_{j=1}^N \hat{\lambda}_j \hat{\m{u}}_j \hat{\m{u}}_j^H = \widehat{\m{U}} \widehat{\m{\Lambda}}\widehat{\m{U}}^H + \widehat{\m{U}}_{\perp} \widehat{\m{\Lambda}}_{\perp}\widehat{\m{U}}_{\perp}^H, }
where $\hat{\cdot}$ denote an estimate of a quantity. The ESPRIT algorithm is implemented by computing the eigenvalues $\lbra{\hat z_k}$ of $\widehat{\m{U}}_1^{\dag}\widehat{\m{U}}_2$. The frequencies are estimated as the angles of $\lbra{\frac{\hat z_k}{\abs{\hat z_k}}}$.

Besides the aforementioned case of $L\rightarrow \infty$, note that if $\sigma\rightarrow 0$ and $\widehat{\m{\Sigma}} = \frac{1}{L} \m{S}\m{S}^H$ is positive definite, then $\widehat{\m{R}}$ has exactly rank $K$ and ESPRIT can exactly localize the frequencies from $\widehat{\m{U}}$.

We summarize the following theorem.
\begin{theorem} The following statements hold true:
\begin{enumerate}
\item Under Assumptions A1--A3, ESPRIT exactly localizes the distinct frequencies $\lbra{f_k}$ in the limiting case of $L\rightarrow \infty$ almost surely if and only if $N\geq K+1$ and $\m{\Sigma}$ is positive definite;
\item Under Assumption A1, ESPRIT exactly localizes the distinct frequencies $\lbra{f_k}$ in the limiting case of $\sigma\rightarrow 0$ if and only if $N\geq K+1$ and $\widehat{\m{\Sigma}}$ is positive definite.
\end{enumerate}\label{thm:ESPRIT}
\end{theorem}

If the number of snapshots $L$ is sufficiently large or the noise variance $\sigma^2$ is sufficiently small, it is natural to expect that $\widehat{\m{U}}$ is a good estimate of $\m{U}$ so that ESPRIT can stably estimate the frequencies.


\subsection{SS-ESPRIT}
A critical assumption for ESPRIT is that the source covariance matrix $\m{\Sigma}$ is positive definite so that the $K$-dimensional range space of $\m{A}$ can be captured with the eigen-decomposition in \eqref{eq:Reigdec2}. SS is a technique to restore the performance of subspace methods in the case when $\m{\Sigma}$ is rank-deficient or ill-conditioned. By SS, the $N$-element physical sensor array is divided into a number of $P=N-M+1$ overlapping $M$-element subarrays and an $M\times M$ smoothed data covariance matrix, denoted by $\m{R}_{\text{SS}}$, is obtained by averaging the covariance matrices of all $P$ subarrays. In particular, the data corresponding to the $p$th subarray, $p=1,\dots,P$, form a submatrix of $\m{Y}$ by collecting $M$ consecutive rows of $\m{Y}$ starting from the $p$th row, denoted by
\equ{\m{Y}_{(p)} = \m{A}_{(p)}\m{S} + \m{E}_{(p)} = \m{A}_M\m{Z}^{p-1}\m{S} + \m{E}_{(p)}, \label{eq:Yp}}
where $\m{A}_{(p)}$ and $\m{E}_{(p)}$ are defined similarly to $\m{Y}_{(p)}$, $\m{A}_M$ is an $M\times K$ Vandermonde matrix that is composed of the first $M$ rows of $\m{A}$, and the identity $\m{A}_{(p)}=\m{A}_M\m{Z}^{p-1}$ holds due to the Vandermonde structure, where $\m{Z} = \diag\sbra{z_1,\dots,z_K}$. Under assumptions A1--A3, the data covariance matrix of the $p$th subarray is thus given by
\equ{\m{R}_p = \bE \frac{1}{L} \m{Y}_{(p)}\m{Y}_{(p)}^H = \m{A}_M\m{Z}^{p-1}\m{\Sigma}\m{Z}^{1-p}\m{A}_M^H + \sigma^2\m{I}. \label{eq:Rp}}
Consequently, the smoothed data covariance matrix is given by
\equ{\m{R}_{\text{SS}} = \frac{1}{P}\sum_{p=1}^P \m{R}_p = \m{A}_M\m{\Sigma}_{\text{SS}} \m{A}_M^H + \sigma^2\m{I}, \label{eq:RSS}}
where
\equ{\m{\Sigma}_{\text{SS}} = \frac{1}{P}\sum_{p=1}^P \m{Z}^{p-1}\m{\Sigma}\m{Z}^{1-p} \label{eq:sSigma}}
is the smoothed source covariance matrix. Evidently, $\m{R}_{\text{SS}}$ has a structure similar to $\m{R}$ and the SS changes $\m{\Sigma}$ to $\m{\Sigma}_{\text{SS}}$ and potentially increases the matrix rank. If $\m{\Sigma}_{\text{SS}}$ has full rank and $M\geq K+1$, then the eigen-decomposition of $\m{R}_{\text{SS}}$ provides the exact signal subspace, based on which the frequencies can be exactly recovered by ESPRIT.

While the aforementioned process is referred to as forward-only SS (FOSS), the technique of forward-backward SS (FBSS) further refines $\m{R}_{\text{SS}}$ to
\equ{\begin{split}\m{R}'_{\text{SS}}
&= \frac{1}{2}\sbra{\m{R}_{\text{SS}} + \m{J} \overline{\m{R}_{\text{SS}}} \m{J}}= \m{A}_M\m{\Sigma}'_{\text{SS}} \m{A}_M^H + \sigma^2\m{I}, \end{split}}
where $\m{J}$ is an $M\times M$ reversal matrix with ones on the anti-diagonal and zeros elsewhere, and
\equ{\m{\Sigma}'_{\text{SS}} = \frac{1}{2}\sbra{\m{\Sigma}_{\text{SS}} + \m{Z}^{1-M}\overline{\m{\Sigma}_{\text{SS}}}\m{Z}^{M-1} }}
is a new smoothed source covariance matrix, where the identity $\m{J}\overline{\m{A}_M} = \m{A}_M\m{Z}^{1-M}$ is used. In practice, both $\m{R}_{\text{SS}}, \m{R}'_{\text{SS}}$ are replaced by their finite-snapshot estimates.

Like Theorem \ref{thm:ESPRIT}, we have the following theorem for SS-ESPRIT.
\begin{theorem} The following statements hold true:
\begin{enumerate}
\item Under Assumptions A1--A3, FOSS-ESPRIT (or FBSS-ESPRIT) exactly localizes the distinct frequencies $\lbra{f_k}$ in the limiting case of $L\rightarrow \infty$ almost surely if and only if $M\geq K+1$ and $\m{\Sigma}_{\text{SS}}$ (or $\m{\Sigma}'_{\text{SS}}$) is positive definite;
\item Under Assumption A1, FOSS-ESPRIT (or FBSS-ESPRIT) exactly localizes the distinct frequencies $\lbra{f_k}$ in the limiting case of $\sigma\rightarrow 0$ if and only if $M\geq K+1$ and $\widehat{\m{\Sigma}}_{\text{SS}}$ (or $\widehat{\m{\Sigma}}'_{\text{SS}}$) is positive definite.
\end{enumerate}\label{thm:SS_ESPRIT}
\end{theorem}

In the case of $L\rightarrow \infty$, the source resolvability of SS-ESPRIT is studied in \cite{yang2019source} from a Hadamard product perspective by writing $\m{\Sigma}_{\text{SS}}$ in \eqref{eq:sSigma} as a Hadamard product, to be specific,
\equ{\begin{split}\m{\Sigma}_{\text{SS}}
&= \frac{1}{P}\sum_{p=1}^P \m{\Sigma}\odot \diag\sbra{\m{Z}^{p-1}}\diag^T\sbra{\m{Z}^{1-p}}\\
&= \frac{1}{P} \m{\Sigma}\odot \sum_{p=1}^P \diag\sbra{\m{Z}^{p-1}}\diag^T\sbra{\m{Z}^{1-p}}\\
&= \frac{1}{P} \m{\Sigma}\odot \overline{\m{A}_P^H\m{A}_P}\\
&= \m{\Sigma}\odot \m{C}_P, \end{split} \label{eq:sSigma2}}
where $\m{C}_P = \frac{1}{P}\overline{\m{A}_P^H\m{A}_P}$ is a $K\times K$ correlation matrix with a unit diagonal and $\m{A}_P$ is $P\times K$ Vandermonde (recall $\m{A}_M$). Consequently, the sources can be resolved with FOSS-ESPRIT if and only if the Hadamard product in \eqref{eq:sSigma2} is positive definite. A similar result holds in the limiting noiseless case by replacing $\m{\Sigma}$ to $\widehat{\m{\Sigma}}$, corresponding to the identifiability problem \cite{yang2019source}.

We also discuss the case of $L=1$. In this case, the matrices $\m{Y}$ and $\m{Y}_{(p)}$ in \eqref{eq:Yp} degenerate into vectors. By making use of \eqref{eq:Rp} and \eqref{eq:RSS}, the smoothed data covariance matrix $\m{R}_{\text{SS}}$ is estimated in practice by
\equ{\begin{split}\widehat{\m{R}}_{\text{SS}}
&= \frac{1}{P}\sum_{p=1}^P \widehat{\m{R}}_p \\
&= \frac{1}{LP}\sum_{p=1}^P \m{Y}_{(p)}\m{Y}_{(p)}^H \\
&= \frac{1}{LP} \mbra{\m{Y}_{(1)},\dots, \m{Y}_{(P)}} \mbra{\m{Y}_{(1)},\dots, \m{Y}_{(P)}}^H,\end{split}}
where it is interesting to note that $\mbra{\m{Y}_{(1)},\dots, \m{Y}_{(P)}}$ is a Hankel matrix. Evidently, the estimated signal and noise subspaces associated with $\widehat{\m{R}}_{\text{SS}}$ can be computed from $\mbra{\m{Y}_{(1)},\dots, \m{Y}_{(P)}}$. Therefore, SS-ESPRIT can be implemented by forming the aforementioned Hankel matrix directly from the observed data. In fact, a similar result holds true in the multisnapshot case (see details in the main context). This perspective on SS-based subspace methods is well-known in the literature on array signal processing (see, e.g., \cite{bresler1986number,thakre2009single}) and has been adopted in \cite{liao2016music,fannjiang2016compressive,aubel2016deterministic,chen2022vectorized}.

\section{Matrix Perturbation Bounds} \label{sec:pertbound}
In this section, we introduce the matrix perturbation theory and present a new matrix perturbation bound for later use. Notations used in this section are self-contained and may be different from other places.

\subsection{Distance and Angles Between Subspaces}
Consider two $r$-dimensional linear subspaces $\cU$, $\widehat{\cU}$ in $\bC^p$ spanned by the columns of $p\times r$ isometric matrices $\m{U}$, $\widehat{\m{U}}$, satisfying $\m{U}^H\m{U} = \widehat{\m{U}}^H \widehat{\m{U}} = \m{I}$. It means that the columns of $\m{U}$ (or $\widehat{\m{U}}$) form an orthonormal basis of $\cU$ (or $\widehat{\cU}$). We will not distinguish a subspace $\cU$ and its matrix representation $\m{U}$ hereafter whenever it is clear from the context. The canonical angles between $\m{U}$ and $\widehat{\m{U}}$ are defined as \equ{\theta_j\sbra{\widehat{\m{U}},\m{U}} = \arccos \sigma_j\sbra{\widehat{\m{U}}^H \m{U}},\quad j=1,\dots,r,}
where $\sigma_j$ denotes the $j$th greatest singular value.
Define matrices
\equ{\Theta = \begin{bmatrix}\theta_1 & & \\ & \ddots & \\ & & \theta_r \end{bmatrix},\quad \sin\Theta = \begin{bmatrix}\sin\theta_1 & & \\ & \ddots & \\ & & \sin\theta_r \end{bmatrix}.}
The distance between $\m{U}$ and $\widehat{\m{U}}$ is defined as
\equ{\text{dist}\sbra{\widehat{\m{U}},\m{U}} = \norm{\sin\Theta\sbra{\widehat{\m{U}},\m{U}}}. \label{eq:distdef}}
It is worth noting that the distance between subspaces can be defined in different but equivalent ways. For example, the same distance as in \eqref{eq:distdef} can be defined as \equ{\text{dist}\sbra{\widehat{\m{U}},\m{U}} = \norm{\cP_{\widehat{\m{U}}} - \cP_{\m{U}} }, \label{eq:distdef2}}
where $\cP_{\m{U}},\cP_{\widehat{\m{U}}}$ denote the orthogonal projections onto $\cU,\widehat{\cU}$ respectively; see, e.g., \cite[Lemma 2.5]{chen2021spectral}.

\subsection{Davis-Kahan and Wedin $\sin\Theta$ Theorems}
The Davis-Kahan $\sin\Theta$ theorem is stated as follows\cite{davis1970rotation}, \cite[Corollary 2.8]{chen2021spectral}.
\begin{theorem} Consider $p\times p$ Hermitian matrices $\m{M}$ and $\widehat{\m{M}}=\m{M}+\m{E}$ that admit the eigen-decompositions:
{\lentwo\equa{ \m{M}
&=& \sum_{j=1}^p \lambda_j \m{u}_j\m{u}_j^H = \begin{bmatrix}\m{U} & \m{U}_{\perp} \end{bmatrix} \begin{bmatrix}\m{\Lambda} & \\ & \m{\Lambda}_{\perp} \end{bmatrix} \begin{bmatrix}\m{U}^H \\ \m{U}_{\perp}^H \end{bmatrix}, \\ \widehat{\m{M}}
&=& \sum_{j=1}^p \widehat{\lambda}_j \widehat{\m{u}}_j\widehat{\m{u}}_j^H = \begin{bmatrix}\widehat{\m{U}} & \widehat{\m{U}}_{\perp} \end{bmatrix} \begin{bmatrix}\widehat{\m{\Lambda}} & \\ & \widehat{\m{\Lambda}}_{\perp} \end{bmatrix} \begin{bmatrix}\widehat{\m{U}}^H \\ \widehat{\m{U}}_{\perp}^H \end{bmatrix},
}}where the eigenvalues $\lbra{\lambda_j}$ and $\lbra{\widehat{\lambda}_j}$ are sorted in descending order, and $\m{U}$, $\m{\Lambda}$, $\widehat{\m{U}}$, $\widehat{\m{\Lambda}}$ are composed of the first $r<p$ eigenvectors or eigenvalues. If $\norm{\m{E}}\leq 0.293\sbra{\lambda_r - \lambda_{r+1}}$, then it holds that
\equ{\text{dist}\sbra{\widehat{\m{U}},\m{U}} \leq \frac{2\norm{\m{E}}}{\lambda_r - \lambda_{r+1}}.} \label{thm:DK}
\end{theorem}

The above result is extended to general matrices by Wedin \cite{wedin1972perturbation},\cite[Theorem 2.9]{chen2021spectral} that is stated in the following theorem.

\begin{theorem} Consider $p\times n$ matrices $\m{M}$ and $\widehat{\m{M}}=\m{M}+\m{E}$ that admit the SVD:
{\lentwo\equa{ \m{M}
&=& \sum_{j=1}^{\min\lbra{p, n}} \sigma_j \m{u}_j\m{v}_j^H = \begin{bmatrix}\m{U} & \m{U}_{\perp} \end{bmatrix} \begin{bmatrix}\m{\Sigma} & \\ & \m{\Sigma}_{\perp} \end{bmatrix} \begin{bmatrix}\m{V}^H \\ \m{V}_{\perp}^H \end{bmatrix}, \\ \widehat{\m{M}}
&=& \sum_{j=1}^{\min\lbra{p, n}} \widehat{\sigma}_j \widehat{\m{u}}_j\widehat{\m{v}}_j^H = \begin{bmatrix}\widehat{\m{U}} & \widehat{\m{U}}_{\perp} \end{bmatrix} \begin{bmatrix}\widehat{\m{\Sigma}} & \\ & \widehat{\m{\Sigma}}_{\perp} \end{bmatrix} \begin{bmatrix}\widehat{\m{V}}^H \\ \widehat{\m{V}}_{\perp}^H \end{bmatrix},
}}where the singular values $\lbra{\sigma_j}$ and $\lbra{\widehat{\sigma}_j}$ are sorted in descending order, and $\m{U}$, $\m{\Sigma}$, $\m{V}$, $\widehat{\m{U}}$, $\widehat{\m{\Sigma}}$, $\widehat{\m{V}}$ are composed of the first $r<\min\lbra{p, n}$ singular vectors or singular values. If $\norm{\m{E}}\leq 0.293\sbra{\sigma_r - \sigma_{r+1}}$, then it holds that
\equ{\max\lbra{\text{dist}\sbra{\widehat{\m{U}},\m{U}}, \text{dist}\sbra{\widehat{\m{V}},\m{V}}} \leq \frac{2\norm{\m{E}}}{\sigma_r - \sigma_{r+1}}.} \label{thm:Wedin}
\end{theorem}

\subsection{A New Matrix Perturbation Bound}
The Wedin's theorem provides a uniform perturbation bound for the left and right singular subspaces. When the dimensions $p$ and $n$ differ significantly, however, the perturbation bound will be suboptimal. Moreover, perturbations are usually caused by random noise in practice. Tighter bounds are expected if such randomness is utilized. We will use the following result in random matrix theory; see, e.g., \cite[Example 6.2]{wainwright2019high}.
\begin{lemma} For $p\times n$ i.i.d.~standard Gaussian random matrix $\m{E}$ and $u>0$, we have
\equ{\norm{\m{E}} \leq \sqrt{p}+\sqrt{n}+u}
with probability at least $1-e^{-\frac{u^2}{2}}$, and
\equ{\norm{\frac{1}{n}\m{E}\m{E}^H - \m{I}} \leq 2\epsilon + \epsilon^2,\quad \epsilon = \sqrt{\frac{p}{n}}+u}
with probability at least $1-2e^{-\frac{n u^2}{2}}$. If $p< n$, then
\equ{\sigma_p\sbra{\m{E}} \geq \sqrt{n}-\sqrt{p}-u}
with probability at least $1-e^{-\frac{u^2}{2}}$.
 \label{lem:Gausnorm}
\end{lemma}

We give a new matrix perturbation bound with i.i.d.~Gaussian noise in the following theorem.

\begin{theorem} Let $\m{M}$ and $\widehat{\m{M}}$ be $p\times n$ matrices given in Theorem \ref{thm:Wedin} and assume that $\m{E}$ is composed of i.i.d.~complex Gaussian entries with zero mean and variance $\sigma^2$. If the upper bound below is less than $0.586$, then we have
\equ{\begin{split}\text{dist}\sbra{\widehat{\m{U}},\m{U}}
\leq \frac{12\sigma\sigma_1\sqrt{p} + 16\sigma^2\max\lbra{\sqrt{pn}, p}} {\sigma_r^2 - \sigma_{r+1}^2} \end{split}\label{eq:newdist}}
with probability at least $1-3e^{-\frac{p}{2}}$. \label{thm:newdist}
\end{theorem}

\begin{proof} See Appendix \ref{app:proof_newdist}.
\end{proof}

Theorem \ref{thm:newdist} is significant if $n\gg p$ and the singular values of $\m{M}$ scales with $\sqrt{n}$. In this case, the upper bound in \eqref{eq:newdist} is proportional to $\sqrt{\frac{p}{n}}$ that vanishes as $n$ approaches infinity, while the Wedin's perturbation bound does not as a contrast.

\begin{remark} When the matrix $\m{M}$ has rank $r$, it is shown in \cite[Theorem 3]{cai2018rate} that
\equ{\begin{split}\bE\,\text{dist}^2\sbra{\widehat{\m{U}},\m{U}}
\leq\frac{Cp\sbra{\sigma^2\sigma_r^2 + \sigma^4n}} {\sigma_r^4}, \end{split}\label{eq:dist3}}
and this upper bound is rate-optimal, where $C$ is a constant. In the case of $n\geq p$, it follows from Theorem \ref{thm:newdist} that
\equ{\begin{split}\text{dist}^2\sbra{\widehat{\m{U}},\m{U}}
\leq\frac{512p\sbra{\sigma^2\sigma_1^2 + \sigma^4n}} {\sigma_r^4} \end{split}\label{eq:dist2}}
with overwhelming probability, showing near optimality of our result. When the matrix $\m{M}$ is random and has rank $r$, we also note a result in \cite[Theorem 3.6]{chen2021spectral} that is similar to ours up to some logarithmic factors.
\end{remark}

\section{Error Bounds for Signal Subspace Estimation} \label{sec:subspaceest}
We quantify the error of signal subspace estimation in ESPRIT and SS-ESPRIT in this section. We consider the ordinary estimation approach for noncoherent sources (without SS) and the SS-based approach, while the latter in the case of FOSS encompasses the former as a special case. Our technique uses the matrix perturbation bound and its proof presented in the previous section.

\subsection{Ordinary Signal Subspace Estimation}
In this case, we have the data model in \eqref{eq:model}. The true signal subspace $\m{U}$ and its estimate $\widehat{\m{U}}$, obtained from $\m{R}$ and $\widehat{\m{R}}$ respectively, are identical to the left singular subspaces of $\m{A}\m{S}$ and $\m{Y}$. The following result is a consequence of Theorem \ref{thm:newdist}.

\begin{theorem} Under Assumption A1, if $\widehat{\m{\Sigma}}$ is positive definite and the upper bound below is less than $0.586$, then it holds for ordinary signal subspace estimation that
\equ{\begin{split}\text{dist}\sbra{\widehat{\m{U}},\m{U}}
&\leq \frac{12\sigma\norm{\m{A}}{\norm{\m{\widehat{\Sigma}}}}^{\frac{1}{2}} \sqrt{\frac{N}{L}} + 16\sigma^2\max\lbra{\sqrt{\frac{N}{L}}, \frac{N}{L}}} {\sigma_K^2\sbra{\m{A}} \lambda_K\sbra{\widehat{\m{\Sigma}}}} \end{split}\label{eq:dist_noSS}}
with probability at least $1-3e^{-\frac{N}{2}}$. \label{thm:dist_noSS}
\end{theorem}
\begin{proof} See Appendix \ref{app:proof_noSS}.
\end{proof}

If we make substitutions $\norm{\m{A}} = \sigma_1\sbra{\m{A}}$ and $\norm{\widehat{\m{\Sigma}}} = \lambda_1\sbra{\widehat{\m{\Sigma}}}$, then the upper bound in \eqref{eq:dist_noSS} can be written as:
\equ{\begin{split}
&\frac{12\sigma\sigma_1\sbra{\m{A}}\lambda_1^{\frac{1}{2}}\sbra{\widehat{\m{\Sigma}}} \sqrt{\frac{N}{L}} + 16\sigma^2\max\lbra{\sqrt{\frac{N}{L}}, \frac{N}{L}}} {\sigma_K^2\sbra{\m{A}} \lambda_K\sbra{\widehat{\m{\Sigma}}}} \\
=& \frac{12\kappa\sbra{\m{A}} \kappa^{\frac{1}{2}}\sbra{\widehat{\m{\Sigma}}} \sqrt{\frac{N}{L}}} {\sqrt{\frac{\sigma_K^2\sbra{\m{A}}\lambda_K\sbra{\widehat{\m{\Sigma}}}}{\sigma^2}}} + \frac{16\max\lbra{\sqrt{\frac{N}{L}}, \frac{N}{L}}} {\frac{\sigma_K^2\sbra{\m{A}}\lambda_K\sbra{\widehat{\m{\Sigma}}}}{\sigma^2}},\notag \end{split}}
where $\kappa\sbra{\m{A}} = \frac{\sigma_1\sbra{\m{A}}}{\sigma_K\sbra{\m{A}}}$, $\kappa\sbra{\widehat{\m{\Sigma}}} = \frac{\lambda_1\sbra{\widehat{\m{\Sigma}}}}{\lambda_K\sbra{\widehat{\m{\Sigma}}}}$ denote the condition numbers and $\frac{\sigma_K^2\sbra{\m{A}}\lambda_K\sbra{\widehat{\m{\Sigma}}}}{\sigma^2}$ can be interpreted as the SNR. It is seen that the upper bound is an increasing function of the condition numbers of $\m{A}$ and $\widehat{\m{\Sigma}}$ and a decreasing function of the SNR. Moreover, the upper bound scales with $\text{SNR}^{-\frac{1}{2}}$ in the high SNR regime and is proportional to $\text{SNR}^{-1}$ in the low SNR regime. Similar interpretations can be made for SS signal subspace estimation to be studied next.

\subsection{FOSS Signal Subspace Estimation}
A crucial assumption in Theorem \ref{thm:dist_noSS} is that $\widehat{\m{\Sigma}}$ is positive definite. If it is not satisfied, then SS can be used to restore the rank of $\widehat{\m{\Sigma}}$ and we consider FOSS in this subsection. In this case, the signal subspace estimate $\widehat{\m{U}}$ is obtained from the smoothed sample covariance matrix $\widehat{\m{R}}_{\text{SS}}$. Recall \eqref{eq:Yp}--\eqref{eq:RSS} and we have
\equ{\begin{split}\widehat{\m{R}}_{\text{SS}}
&= \frac{1}{P}\sum_{p=1}^P \widehat{\m{R}}_p\\
&=\frac{1}{LP}\sum_{p=1}^P \m{Y}_{(p)}\m{Y}_{(p)}^H\\
&=\frac{1}{LP} \mbra{\m{Y}_{(1)},\; \dots,\;\m{Y}_{(P)}}\mbra{\m{Y}_{(1)},\; \dots,\;\m{Y}_{(P)}}^H. \end{split} \label{eq:RSShat}}
Consequently, $\widehat{\m{U}}$ is the left singular subspace of the matrix
\equ{\begin{split}\m{Y}_{\text{SS}}
&= \mbra{\m{Y}_{(1)},\; \dots,\;\m{Y}_{(P)}}\\
&= \mbra{\m{A}_M\m{S},\;\dots,\; \m{A}_M\m{Z}^{P-1}\m{S}} + \mbra{\m{E}_{(1)},\; \dots,\;\m{E}_{(P)}}. \end{split} \label{eq:YSS}}
Moreover, $\m{U}$ is the left singular subspace of
\equ{\begin{split}\mbra{\m{A}\m{S}}_{\text{SS}}
&=\mbra{\m{A}_M\m{S},\;\dots,\; \m{A}_M\m{Z}^{P-1}\m{S}} \\
&= \m{A}_M\mbra{\m{S},\;\dots,\; \m{Z}^{P-1}\m{S}} \end{split}}
if $\mbra{\m{A}\m{S}}_{\text{SS}}$ has rank $K$.
Therefore, we have the perturbation model
\equ{\m{Y}_{\text{SS}} = \mbra{\m{A}\m{S}}_{\text{SS}} + \m{E}_{\text{SS}}, \label{eq:YSSmodel}}
where $\m{E}_{\text{SS}}$ is defined similarly as $\m{Y}_{\text{SS}}$. The new model \eqref{eq:YSSmodel} is similar to \eqref{eq:model}, but $\m{E}_{\text{SS}}$ is no longer i.i.d.~Gaussian and thus Theorem \ref{thm:newdist} cannot be applied. Instead, we use proof techniques similar to those for Theorem \ref{thm:newdist} to derive the following result.

\begin{theorem} Under Assumption A1, if $\widehat{\m{\Sigma}}_{\text{SS}} = \widehat{\m{\Sigma}}\odot\m{C}_P$ is positive definite and the upper bound below is less than $0.586$, then it holds for FOSS signal subspace estimation that
\equ{\begin{split}
&\text{dist}\sbra{\widehat{\m{U}},\m{U}}\\
&\leq\frac{12\sigma\norm{\m{A}_M} \norm{\widehat{\m{\Sigma}}}^{\frac{1}{2}}\sqrt{\frac{M}{L}} + 16\sigma^2 \max\lbra{\sqrt{\frac{M}{L}}, \frac{M}{L}}} {\sigma_K^2\sbra{\m{A}_M} \lambda_K\sbra{\widehat{\m{\Sigma}}_{\text{SS}}}} \end{split}\label{eq:dist_SS}}
with probability at least $1-3Pe^{-\frac{M}{2}}$.  \label{thm:dist_SS}
\end{theorem}
\begin{proof} See Appendix \ref{app:proof_SS}.
\end{proof}

Theorem \ref{thm:dist_SS} degenerates into Theorem \ref{thm:dist_noSS} in the case of $P=1$ in which no SS is used and we have $M=N$ and $\widehat{\m{\Sigma}}_{\text{SS}} = \widehat{\m{\Sigma}}$.

\subsection{FBSS Signal Subspace Estimation}
Similarly to FOSS, we have the following result for FBSS.
\begin{theorem} Under Assumption A1, if $\widehat{\m{\Sigma}}'_{\text{SS}} = \frac{1}{2}\sbra{\widehat{\m{\Sigma}}_{\text{SS}} +\m{Z}^{1-M}\overline{\widehat{\m{\Sigma}}_{\text{SS}}}\m{Z}^{M-1}}$ is positive definite and the upper bound below is less than $0.586$, then it holds for FBSS signal subspace estimation that
\equ{\begin{split}
&\text{dist}\sbra{\widehat{\m{U}},\m{U}}\\
&\leq \frac{12\sigma\norm{\m{A}_M} \norm{\widehat{\m{\Sigma}}}^{\frac{1}{2}}\sqrt{\frac{M}{L}} + 16\sigma^2 \max\lbra{\sqrt{\frac{M}{L}}, \frac{M}{L}}} {\sigma_K^2\sbra{\m{A}_M} \lambda_K\sbra{\widehat{\m{\Sigma}}'_{\text{SS}}}} \end{split}\label{eq:dist_FBSS}}
with probability at least $1-3Pe^{-\frac{M}{2}}$.  \label{thm:dist_FBSS}
\end{theorem}
\begin{proof} See Appendix \ref{app:proof_FBSS}.
\end{proof}

The error bound for FBSS in Theorem \ref{thm:dist_FBSS} is stronger than that for FOSS in Theorem \ref{thm:dist_SS}. To see this, note that $\m{Z}^{1-M}\overline{\widehat{\m{\Sigma}}_{\text{SS}}}\m{Z}^{M-1}$, $\overline{\widehat{\m{\Sigma}}_{\text{SS}}}$ and $\widehat{\m{\Sigma}}_{\text{SS}}$ share the same eigenvalues. Consequently,
\equ{\begin{split}\lambda_K\sbra{\widehat{\m{\Sigma}}'_{\text{SS}}}
&\geq \frac{1}{2}\sbra{\lambda_K\sbra{\widehat{\m{\Sigma}}_{\text{SS}}} + \lambda_K\sbra{\m{Z}^{1-M}\overline{\widehat{\m{\Sigma}}_{\text{SS}}}\m{Z}^{M-1}}} \\
&= \lambda_K\sbra{\widehat{\m{\Sigma}}_{\text{SS}}}. \end{split} \label{eq:eigineq}}
If $\lambda_K\sbra{\widehat{\m{\Sigma}}_{\text{SS}}}>0$, i.e., $\widehat{\m{\Sigma}}_{\text{SS}}$ is positive definite, then it follows from \eqref{eq:eigineq} that $\widehat{\m{\Sigma}}_{\text{SS}}'$ is positive definite and the upper bound in \eqref{eq:dist_FBSS} is no greater than that in \eqref{eq:dist_SS}. It is also possible that $\widehat{\m{\Sigma}}_{\text{SS}}$ is singular, which means that the assumption in Theorem \ref{thm:dist_SS} is not satisfied, but $\widehat{\m{\Sigma}}_{\text{SS}}'$ is positive definite so that Theorem \ref{thm:dist_FBSS} is still applicable.

\section{Stability and Resolution of ESPRIT and SS-ESPRIT} \label{sec:bound}

%
%
%
%
%
%

\subsection{Stability}
For the frequency set $\cT=\lbra{f_k}_{k=1}^K$ and its estimate $\widehat{\cT}=\lbra{\widehat f_k}_{k=1}^K$, define their matched (wrapped-around) distance as \cite{li2020super}:
\equ{\begin{split}
&\text{md}\sbra{\widehat{\cT},\cT} \\
&= \min_{\psi}\max_{k}\min\lbra{\abs{\widehat f_{\psi(k)} - f_k}, 1-\abs{\widehat f_{\psi(k)} - f_k}},\end{split}}
where $\psi$ is a permutation on $\lbra{1,\dots,K}$. By definition, the matched distance measures the maximum absolute error of frequency estimation on the unit circle. The following result is the key to the analysis of single-snapshot ESPRIT in \cite{li2020super}, which is summarized in \cite[Lemma A.7]{li2021stability} as a combination of Lemmas 2, 3 and 6 in \cite{li2020super}.
\begin{lemma} For the ordinary signal subspace estimate $\widehat{\m{U}}$, if $N\geq K+1$, then it holds for ESPRIT that
\equ{\text{md}\sbra{\widehat{\cT},\cT} \leq \frac{2^{2K+4}K^{3/2}\sqrt{N}} {\sigma_K\sbra{\m{A}}} \text{dist}\sbra{\widehat{\m{U}},\m{U}}. \label{eq:mdlem}} \label{lem:mdlem}
\end{lemma}

Combining Lemma \ref{lem:mdlem} and Theorems \ref{thm:dist_noSS}, \ref{thm:dist_SS} and \ref{thm:dist_FBSS} immediately results in the following theorem.



\begin{theorem} Under Assumption A1, the following statements hold true:
\begin{enumerate}
\item If $N\geq K+1$ and $\widehat{\m{\Sigma}}$ is positive definite, then it holds for ESPRIT that
\equ{\begin{split}
&\text{md}\sbra{\widehat{\cT},\cT} \leq \min \left\{1,\; 2^{2K+4}K^{3/2}\sqrt{N} \phantom{\frac{{\norm{\m{\widehat{\Sigma}}}}^{\frac{1}{2}}} { \lambda_K\sbra{\widehat{\m{\Sigma}}}}} \right. \\
&\;\times \left.\frac{12\sigma\norm{\m{A}}{\norm{\m{\widehat{\Sigma}}}}^{\frac{1}{2}} \sqrt{\frac{N}{L}} + 16\sigma^2\max\lbra{\sqrt{\frac{N}{L}}, \frac{N}{L}}} {\sigma_K^3\sbra{\m{A}} \lambda_K\sbra{\widehat{\m{\Sigma}}}} \right\} \end{split}\label{eq:md_noSS1}}
with probability at least $1-3e^{-\frac{N}{2}}$;
\item If $M\geq K+1$ and $\widehat{\m{\Sigma}}_{\text{SS}}$ is positive definite, then it holds for FOSS-ESPRIT that
\equ{\begin{split}
&\text{md}\sbra{\widehat{\cT},\cT} \leq \min \left\{1,\; 2^{2K+4}K^{3/2}\sqrt{M} \phantom{\frac{{\norm{\m{\widehat{\Sigma}}}}^{\frac{1}{2}}} { \lambda_K\sbra{\widehat{\m{\Sigma}}}}} \right. \\
&\;\times \left.\frac{12\sigma\norm{\m{A}_M}{\norm{\m{\widehat{\Sigma}}}}^{\frac{1}{2}} \sqrt{\frac{M}{L}} + 16\sigma^2\max\lbra{\sqrt{\frac{M}{L}}, \frac{M}{L}}} {\sigma_K^3\sbra{\m{A}_M} \lambda_K\sbra{\widehat{\m{\Sigma}}_{\text{SS}}}} \right\} \end{split}\label{eq:md_SS1}}
with probability at least $1-3Pe^{-\frac{M}{2}}$;
\item If $M\geq K+1$ and $\widehat{\m{\Sigma}}'_{\text{SS}}$ is positive definite, then it holds for FBSS-ESPRIT that
\equ{\begin{split}
&\text{md}\sbra{\widehat{\cT},\cT} \leq \min \left\{1,\; 2^{2K+4}K^{3/2}\sqrt{M} \phantom{\frac{{\norm{\m{\widehat{\Sigma}}}}^{\frac{1}{2}}} { \lambda_K\sbra{\widehat{\m{\Sigma}}}}} \right. \\
&\;\times \left.\frac{12\sigma\norm{\m{A}_M}{\norm{\m{\widehat{\Sigma}}}}^{\frac{1}{2}} \sqrt{\frac{M}{L}} + 16\sigma^2\max\lbra{\sqrt{\frac{M}{L}}, \frac{M}{L}}} {\sigma_K^3\sbra{\m{A}_M} \lambda_K\sbra{\widehat{\m{\Sigma}}'_{\text{SS}}}} \right\} \end{split}\label{eq:md_FBSS1}}
with probability at least $1-3Pe^{-\frac{M}{2}}$.
\end{enumerate} \label{thm:mdthm1}
\end{theorem}
\begin{proof} Inserting \eqref{eq:dist_noSS}, \eqref{eq:dist_SS} and \eqref{eq:dist_FBSS}, respectively, into \eqref{eq:mdlem} completes the proof. Note that the unit upper bound naturally holds. The condition that the upper bounds in \eqref{eq:dist_noSS}, \eqref{eq:dist_SS} and \eqref{eq:dist_FBSS} are less than $0.586$ is removed since the upper bound in \eqref{eq:mdlem} makes sense only if
\equ{\begin{split}\text{dist}\sbra{\widehat{\m{U}},\m{U}}
&\leq \frac{\sigma_K\sbra{\m{A}_N}}{2^{2K+4}K^{3/2}\sqrt{N}} \\
&\leq \frac{\sqrt{\frac{1}{K}\frobn{\m{A}}^2}}{2^{2K+4}K^{3/2}\sqrt{N}} \\
&= \frac{1}{2^{2K+4}K^{3/2}} \\
&< 0.586. \end{split}}
\end{proof}

It follows from Theorem \ref{thm:mdthm1} that ESPRIT and SS-ESPRIT can stably estimate the frequencies for any fixed $L$ under mild conditions provided that $\sigma$ is sufficiently small. In fact, the conditions in Theorem \ref{thm:mdthm1} are necessary for ESPRIT or SS-ESPRIT to work even in the limiting noiseless case according to Theorems \ref{thm:ESPRIT} and \ref{thm:SS_ESPRIT}. For light noise with $\sigma \ll \norm{\m{A}_M} \norm{\widehat{\m{\Sigma}}}^{\frac{1}{2}}$, the estimation errors of ESPRIT and SS-ESPRIT scale linearly with the noise level $\sigma$.

\begin{remark} While this paper is focused on the multiple-snapshot case, our result on FOSS-ESPRIT in the single-snapshot case is suboptimal, as compared to \cite{li2020super}, due to the suboptimal estimate of the noise perturbation in this case (see Theorem \ref{thm:dist_SS} and its proof). But still, the aforementioned linear scaling behavior with respect to the noise level $\sigma$ is consistent with that in \cite{li2020super}. We note that a similar scaling behavior is also shown in \cite{aubel2016deterministic} for the single-snapshot case with a different nonasymptotic analysis. A detailed comparison of the error bounds in \cite{li2020super} and \cite{aubel2016deterministic} is nontrivial and is beyond the scope of this paper.
\end{remark}

To show how the frequency estimation error scales with $L$, we give the following theorem.

\begin{theorem} Under Assumptions A1--A3, the following statements hold true:
\begin{enumerate}
\item If $N\geq K+1$, $\m{\Sigma}$ is positive definite and $L\geq \max\lbra{N, 16K}$, then it holds for ESPRIT that
\equ{\begin{split}
&\text{md}\sbra{\widehat{\cT},\cT} \leq \min \left\{1,\;  \phantom{\frac{\max\lbra{\sigma\norm{\m{A}}\norm{\m{\Sigma}}^{\frac{1}{2}}, \sigma^2} } {\sqrt{L}}} \right. \\
&\quad \left.\frac{72 \cdot 2^{2K+5}K^{3/2}N} {\sigma_K^3\sbra{\m{A}} \lambda_K\sbra{\m{\Sigma}}}\cdot\frac{\max\lbra{\sigma\norm{\m{A}}\norm{\m{\Sigma}}^{\frac{1}{2}}, \sigma^2} } {\sqrt{L}} \right\} \end{split}\label{eq:mdthm_noSS}}
with probability at least $1-3e^{-\frac{N}{2}}-2e^{-\frac{L}{32}}$;
\item If $M\geq K+1$, $\m{\Sigma}_{\text{SS}}$ is positive definite and $L\geq \max\lbra{M, 16\rank\sbra{\m{\Sigma}}}$, then it holds for FOSS-ESPRIT that
\equ{\begin{split}
&\text{md}\sbra{\widehat{\cT},\cT} \leq \min \left\{1,\;  \phantom{\frac{\max\lbra{\sigma\norm{\m{A}}\norm{\m{\Sigma}}^{\frac{1}{2}}, \sigma^2} } {\sqrt{L}}} \right. \\
&\quad \left.\frac{72 \cdot 2^{2K+5}K^{3/2}M} {\sigma_K^3\sbra{\m{A}_M} \lambda_K\sbra{\m{\Sigma}_{\text{SS}}}}\cdot \frac{\max\lbra{\sigma\norm{\m{A}_M}\norm{\m{\Sigma}}^{\frac{1}{2}}, \sigma^2} } {\sqrt{L}} \right\} \end{split}\label{eq:mdthm_SS}}
with probability at least $1-3Pe^{-\frac{M}{2}}-2e^{-\frac{L}{32}}$;
\item If $M\geq K+1$, $\m{\Sigma}'_{\text{SS}}$ is positive definite and $L\geq \max\lbra{M, 16\rank\sbra{\m{\Sigma}}}$, then it holds for FBSS-ESPRIT that
\equ{\begin{split}
&\text{md}\sbra{\widehat{\cT},\cT} \leq \min \left\{1,\;  \phantom{\frac{\max\lbra{\sigma\norm{\m{A}}\norm{\m{\Sigma}}^{\frac{1}{2}}, \sigma^2} } {\sqrt{L}}} \right. \\
&\quad \left.\frac{72 \cdot 2^{2K+5}K^{3/2}M} {\sigma_K^3\sbra{\m{A}_M} \lambda_K\sbra{\m{\Sigma}'_{\text{SS}}}} \cdot \frac{\max\lbra{\sigma\norm{\m{A}_M}\norm{\m{\Sigma}}^{\frac{1}{2}}, \sigma^2} } {\sqrt{L}} \right\} \end{split}\label{eq:mdthm_FBSS}}
with probability at least $1-3Pe^{-\frac{M}{2}}-2e^{-\frac{L}{32}}$.
\end{enumerate} \label{thm:mdthm}
\end{theorem}

\begin{proof} See Appendix \ref{app:proof_mdthm}.
\end{proof}

It follows from Theorem \ref{thm:mdthm} that ESPRIT and SS-ESPRIT can stably estimate the frequencies for any fixed noise level $\sigma$ under mild conditions provided that $L$ is sufficient large. These conditions are exactly those in Theorems \ref{thm:ESPRIT} and \ref{thm:SS_ESPRIT} which are necessary to guarantee exact frequency localization of ESPRIT and SS-ESPRIT with infinitely many snapshots. Combining Theorems \ref{thm:mdthm1} and \ref{thm:mdthm} (and comparing with Theorems \ref{thm:ESPRIT} and \ref{thm:SS_ESPRIT}), we conclude that ESPRIT and SS-ESPRIT can stably estimate the frequencies if
\equ{\frac{\max\lbra{\sigma, \sigma^2}}{\sqrt{L}}}
is small, conditioning on that the algorithms succeed to locate the true frequencies in the limiting case of $\sigma\rightarrow 0$ or $L\rightarrow \infty$. Therefore, for both ESPRIT and SS-ESPRIT there is no substantial gap either between the noiseless and the noisy cases or between the infinite-snapshot and finite-snapshot cases.

\subsection{Resolution}
\begin{definition} We say that an algorithm achieves resolution $\Delta$ if it resolves a set of frequencies $\cT$, which has minimum separation $\Delta = \min_{p\neq q} \min\lbra{\abs{f_p-f_q}, 1-\abs{f_p-f_q}}$, with precision
\equ{\text{md}\sbra{\widehat{\cT},\cT}< \frac{\Delta}{2}.}
\end{definition}

The following result is a corollary to Theorem \ref{thm:mdthm} and shows that ESPRIT and SS-ESPRIT can achieve arbitrarily high resolution given sufficiently large $L$.

\begin{corollary} Under Assumptions A1--A3, the following statements hold true:
\begin{enumerate}
\item If $N\geq K+1$, $\m{\Sigma}$ is positive definite and
\equ{\begin{split}
&L > \max\left\{N, \;16K,\;\phantom{\frac{\max\lbra{\norm{\m{A}}^2}} {\sigma_K^6\sbra{\m{A}}}} \right.\\
&\quad \left.\frac{72^2 \cdot 2^{4K+12}K^{3}N^2 \max\lbra{\sigma^2\norm{\m{A}}^2\norm{\m{\Sigma}}, \sigma^4} } {\sigma_K^6\sbra{\m{A}} \lambda_K^2\sbra{\m{\Sigma}}\Delta^2}\right\}, \end{split} \label{eq:resthm_noSS}}
then ESPRIT is guaranteed to achieve resolution $\Delta$ with probability at least $1-3e^{-\frac{N}{2}}-2e^{-\frac{L}{32}}$;
\item If $M\geq K+1$, $\m{\Sigma}_{\text{SS}}$ is positive definite and
\equ{\begin{split}
&L > \max\left\{M, \;16\rank\sbra{\m{\Sigma}},\;\phantom{\frac{\max\lbra{\norm{\m{A}}^2}} {\sigma_K^6\sbra{\m{A}}}} \right.\\
&\quad \left.\frac{72^2 \cdot 2^{4K+12}K^{3}M^2 \max\lbra{\sigma^2\norm{\m{A}_M}^2\norm{\m{\Sigma}}, \sigma^4 }} {\sigma_K^6\sbra{\m{A}_M} \lambda_K^2\sbra{\m{\Sigma}_{\text{SS}}}\Delta^2}\right\}, \end{split} \label{eq:resthm_SS}}
then FOSS-ESPRIT is guaranteed to achieve resolution $\Delta$ with probability at least $1-3Pe^{-\frac{M}{2}}-2e^{-\frac{L}{32}}$;
\item If $M\geq K+1$, $\m{\Sigma}'_{\text{SS}}$ is positive definite and
\equ{\begin{split}
&L > \max\left\{M, \;16\rank\sbra{\m{\Sigma}},\;\phantom{\frac{\max\lbra{\norm{\m{A}}^2}} {\sigma_K^6\sbra{\m{A}}}} \right.\\
&\quad \left.\frac{72^2 \cdot 2^{4K+12}K^{3}M^2 \max\lbra{\sigma^2\norm{\m{A}_M}^2\norm{\m{\Sigma}}, \sigma^4 }} {\sigma_K^6\sbra{\m{A}_M} \lambda_K^2\sbra{\m{\Sigma}'_{\text{SS}}}\Delta^2}\right\}, \end{split} \label{eq:resthm_FBSS}}
then FBSS-ESPRIT is guaranteed to achieve resolution $\Delta$ with probability at least $1-3Pe^{-\frac{M}{2}}-2e^{-\frac{L}{32}}$.
\end{enumerate} \label{thm:resthm}
\end{corollary}
\begin{proof} Letting the upper bounds in \eqref{eq:mdthm_noSS}, \eqref{eq:mdthm_SS} and \eqref{eq:mdthm_FBSS} be less than $\frac{\Delta}{2}$ proves the corollary.
\end{proof}

%
%

%

\section{Positive-Definiteness and Minimum Eigenvalue of Hadamard Products} \label{sec:hadamard}
Positive-definiteness and the minimum (i.e., the $K$-th) eigenvalue of the Hadamard product $\m{\Sigma}\odot \m{C}_P$ (or $\widehat{\m{\Sigma}}\odot \m{C}_P$) are involved in theorems regarding SS-ESPRIT. To well understand these theorems, we study how they depends on the matrix factors $\m{\Sigma}$ and $\m{C}_P=\frac{1}{P}\overline{\m{A}_P^H\m{A}_P}$ in this section. Note that both $\m{\Sigma},\m{C}_P$ are positive semidefinite and have positive diagonals. The matrix $\m{\Sigma}$ is singular in presence of coherent sources, and $\m{C}_P$ is singular if $P<K$ since $\rank\sbra{\m{C}_P} = \rank\sbra{\m{A}_P}=\min\lbra{P, K}$, where $P+M = N+1$.

\subsection{Positive-Definiteness of $\m{\Sigma}\odot \m{C}_P$}
The Hadamard product of positive semidefinite matrices is concerned by the classical Schur product theorem \cite[Theorem VII]{schur1911bemerkungen}, of which an inclusive statement is given below.\footnote{It is interesting to note that the third statement in Theorem \ref{thm:schur} was usually excluded from the Schur product theorem (see, e.g., \cite[Theorem 7.5.3]{horn2012matrix}, \cite[Fact 8.21.12]{bernstein2009matrix} and \cite{yang2019source}), though it is a direct consequence of \eqref{eq:hpeig} that is the main result of \cite[Theorem VII]{schur1911bemerkungen}.}

\begin{theorem} Every eigenvalue of the Hadamard product of positive semidefinite matrices $\m{B},\m{C}$ satisfies
\equ{\lambda_{\text{max}}\sbra{\m{B}} \max_j C_{jj} \geq \lambda_{l}\sbra{\m{B}\odot\m{C}} \geq \lambda_{\text{min}}\sbra{\m{B}} \min_j C_{jj}, \label{eq:hpeig}}
where the positions of $\m{B},\m{C}$ can be swapped since $\m{B}\odot\m{C} = \m{C}\odot\m{B}$. Three direct consequences are the following:
\begin{enumerate}
\item $\m{B}\odot\m{C}\geq \m{0}$ if $\m{B}\geq \m{0}$ and $\m{C}\geq \m{0}$;
\item $\m{B}\odot\m{C}>\m{0}$ if $\m{B}>\m{0}$ and $\m{C}>\m{0}$;
\item $\m{B}\odot\m{C}>\m{0}$ if one of $\m{B}, \m{C}\geq\m{0}$ is positive definite and the other has a positive diagonal.
\end{enumerate}
\label{thm:schur}
\end{theorem}

We next revisit our previous results in \cite{yang2019source} in which the Hadamard product is used to study exact frequency localization of ESPRIT with infinitely many snapshots. According to Theorem \ref{thm:schur}, $\m{\Sigma}\odot \m{C}_P$ is guaranteed to be positive definite if either $\m{\Sigma}$ or $\m{C}_P$ is positive definite. The former case means that all sources are noncoherent, resulting in $P\geq 1$ and the array size $N = M+P-1\geq K+1$ (note that $M\geq K+1$). In the latter case, we have $P\geq K$, leading to an array size $N \geq 2K$, which recovers the result in \cite{shan1985spatial}.

When both $\m{\Sigma}, \m{C}_P$ are singular, sufficient conditions are provided in \cite{yang2019source} to ensure that $\m{\Sigma}\odot \m{C}_P$ is positive definite by extending the Schur product theorem to the case when both matrix factors are singular. Assume that the $K$ sources are divided into $G\leq K$ coherent groups and the $j$th group is composed of $g_j$ coherent sources, where $K = \sum_{j=1}^{G} g_j$ and $g_1\geq \dots\geq g_G\geq 1$. Without loss of generality, we assume that the sources are indexed according to the above coherency structure so that $\m{\Sigma}$ admits the factorization
\equ{\m{\Sigma} = \diag\sbra{\m{v}_1,\dots,\m{v}_G}\check{\m{\Sigma}} \diag^H\sbra{\m{v}_1,\dots,\m{v}_G}, \label{eq:Sigmadec}}
where $\m{v}_j$ is a $g_j\times 1$ vector with nonzero entries and $\norm{\m{v}_j}=1$ for $j=1,\dots,G$ and $\check{\m{\Sigma}}$ is $G\times G$ positive semidefinite with $\rank\sbra{\check{\m{\Sigma}}} = \rank\sbra{\m{\Sigma}}$. It is shown in \cite{yang2019source} that if
\equ{P\geq \sum_{j=1}^{G-\rank\sbra{\check{\m{\Sigma}}}+1} g_j,}
which yields that
\equ{N\geq K + \sum_{j=1}^{G-\rank\sbra{\check{\m{\Sigma}}}+1} g_j,}
then $\m{\Sigma}\odot \m{C}_P$ is guaranteed to be positive definite. This result remains true when partial or even no knowledge of the coherence structure is available. For example, when no coherence structure is known, we can choose $G=K$ and $g_1= \dots= g_G= 1$, which results in $N\geq 2K-\rank\sbra{\m{\Sigma}}+1$ and recovers the result in \cite{bresler1986number}. If $\rank\sbra{\check{\m{\Sigma}}}=G$, then $N\geq K+g_1$, which recovers the result in \cite{shan1987smoothed}.

\subsection{Minimum Eigenvalue of $\m{\Sigma}\odot \m{C}_P$}
In the case of noncoherent sources with positive definite $\m{\Sigma}$, it follows from \eqref{eq:hpeig} that
\equ{\lambda_{\text{min}}\sbra{\m{\Sigma}\odot \m{C}_P} \geq \lambda_{\text{min}}\sbra{\m{\Sigma}} \label{eq:HPlb1}}
by recalling that $\m{C}_P$ is a correlation matrix with a unit diagonal. If $P\geq K$ so that $\m{C}_P$ is positive definite, then we have
\equ{\begin{split}\lambda_{\text{min}}\sbra{\m{\Sigma}\odot \m{C}_P}
&= \lambda_{\text{min}}\sbra{\m{C}_P\odot\m{\Sigma}}\\
&\geq \lambda_{\text{min}}\sbra{\m{C}_P}\cdot \min_j \Sigma_{jj} \\
&= \frac{1}{P}\sigma_K^2\sbra{\m{A}_P}\cdot \min_j \Sigma_{jj}, \end{split}\label{eq:HPlb2}}
where $\min_j\Sigma_{jj}$ represents the smallest source power.

The challenge arises in the case when both $\m{\Sigma}, \m{C}_P$ are singular. In this case the lower bound in \eqref{eq:hpeig} is trivially zero. Sufficient conditions are provided in \cite{yang2019source} to ensure positive-definiteness of $\m{\Sigma}\odot \m{C}_P$, but its minimum eigenvalue is not explicitly measured. We show the following result in this paper.

\begin{proposition} Given $\m{\Sigma}$ in \eqref{eq:Sigmadec} and partitioning $\m{A}_P$ into a $1\times G$ block matrix according to the coherence structure $\lbra{g_j}_{j=1}^G$ such that $\m{A}_P=\mbra{\m{A}_P^{\sbra{1}},\dots,\m{A}_P^{\sbra{G}}}$, where $\m{A}_P^{\sbra{j}}$ is a $P\times g_j$ Vandermonde matrix, we have
\equ{\lambda_{\text{min}}\sbra{\m{\Sigma}\odot \m{C}_P} \geq \frac{1}{P}\lambda_{G}\sbra{\m{\Sigma}} \min_j \lbra{\sigma_{g_j}^2\sbra{\m{A}_P^{\sbra{j}}} \min_l \abs{v_{jl}}^2}, \label{eq:newHPlb}}
where $\lambda_{G}\sbra{\m{\Sigma}} =\lambda_{\text{min}}\sbra{\check {\m{\Sigma}}}$. \label{prop:newHPlb}
\end{proposition}
\begin{proof} See Appendix \ref{app:proof_newHPlb}.
\end{proof}

By Proposition \ref{prop:newHPlb}, the minimum eigenvalue of $\m{\Sigma}\odot \m{C}_P$ depends on the smallest positive eigenvalue of $\m{\Sigma}$, diversity of source powers in coherent groups, and the minimum singular value of the steering matrix $\lbra{\m{A}_P^{\sbra{j}}}$ regarding each coherent group. The lower bound in \eqref{eq:newHPlb} is strictly positive if $\check{\m{\Sigma}}>0$ and $P\geq g_1$, yielding $N\geq K+g_1$. It recovers \eqref{eq:HPlb1} if $\m{\Sigma}$ is positive definite. For general positive semidefinite and singular matrices $\m{B},\m{C}$, it remains an open problem to quantify the minimum eigenvalue of $\m{B}\odot\m{C}$ provided that the Hadamard product is positive definite.

The minimum singular value of a tall Vandermonde matrix, say the $N\times K$ matrix $\m{A}$, plays an important role in the lower bounds in \eqref{eq:HPlb2} and \eqref{eq:newHPlb}. This topic has recently been extensively studied, see \cite{moitra2015super,pan2016bad,liao2016music,batenkov2021single} and references therein. It is shown that $\sigma_K\sbra{\m{A}}$ mainly depends on the separations between adjacent frequencies composing $\m{A}$, especially on the minimum separation. When proper separations are assumed, $\sigma_K\sbra{\m{A}}$ scales with $\sqrt{N}$.

\section{Numerical Results} \label{sec:simulation}
\begin{figure}[htbp] \centerline{\includegraphics[width=3.5in]{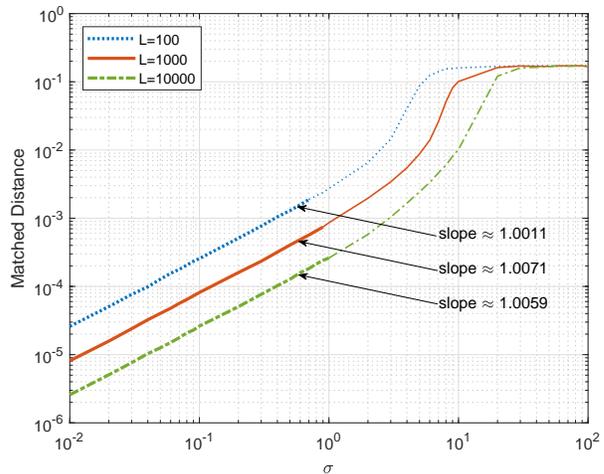}}
\caption{Results of matched distance of frequency estimation using ESPRIT versus the noise level $\sigma$.}
	\label{fig:esprit_noise}
\end{figure}

We validate our theoretical analysis with numerical results in this section. In {\em Experiment 1}, we consider $K=3$ noncoherent sources with the frequency set $\cT = \lbra{0.1,0.5, 0.8}$ that corresponds to the set of DOAs $\lbra{11.54^\circ, -90^\circ, -23.58^\circ}$. The source signals are i.i.d.~and generated from a standard complex Gaussian distribution. We use $N=10$ samples per snapshot and $L$ snapshots that are corrupted by i.i.d.~complex Gaussian noise with zero mean and variance $\sigma^2$, where $L$ takes value in $\lbra{10^2,10^3,10^4}$ and $\sigma$ ranges from $10^{-2}$ to $10^2$. The ESPRIT algorithm is used to estimate the frequencies. For each combination $\sbra{L,\sigma}$, a number of 1000 Monte Carlo runs are carried out and the matched distance of frequency estimation is obtained by averaging the results. The numerical results are presented in Fig.~\ref{fig:esprit_noise}. It is seen that for each $L$, the matched distance curve is straight with a unit slope in a broad range of noise level, implying that the matched distance scales with $\sigma$ in this range, as predicted by Theorems \ref{thm:mdthm1} and \ref{thm:mdthm}. The frequency estimation error propagates faster as $\sigma$ increases beyond this range until the algorithm fails to localize the frequencies.

\begin{figure}[htbp] \centerline{\includegraphics[width=3.5in]{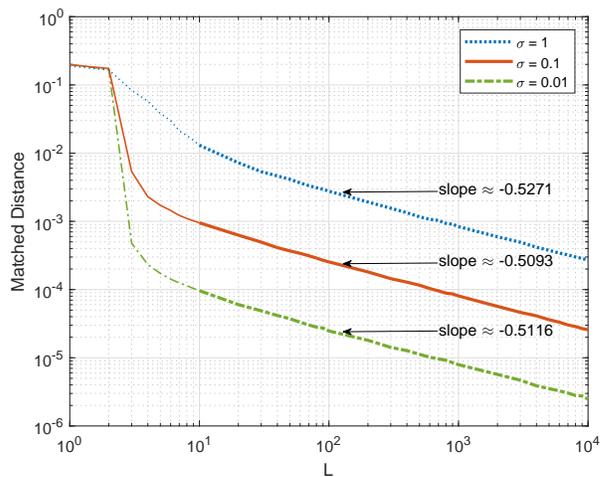}}
\caption{Results of matched distance of frequency estimation using ESPRIT versus the number of snapshots $L$.}
	\label{fig:esprit_L}
\end{figure}

In {\em Experiment 2}, we repeat {\em Experiment 1} by fixing the noise level $\sigma\in\lbra{0.01, 0.1, 1}$ and varying $L$ from $1$ to $10^4$. Our numerical results are presented in Fig.~\ref{fig:esprit_L}. It is seen that the matched distance curves are approximately straight with a slope of about $-0.5$ as $L\geq 10$, implying that the frequency estimation error of ESPRIT scales with $\frac{1}{\sqrt{L}}$, as predicted by Theorems \ref{thm:mdthm1} and \ref{thm:mdthm}. Note that ESPRIT can stably estimate the frequencies only if $L\geq K =3$.

\begin{figure}[htbp]
\subfigure[]{
\label{fig:res_L}
\centering
\includegraphics[width=3.5in]{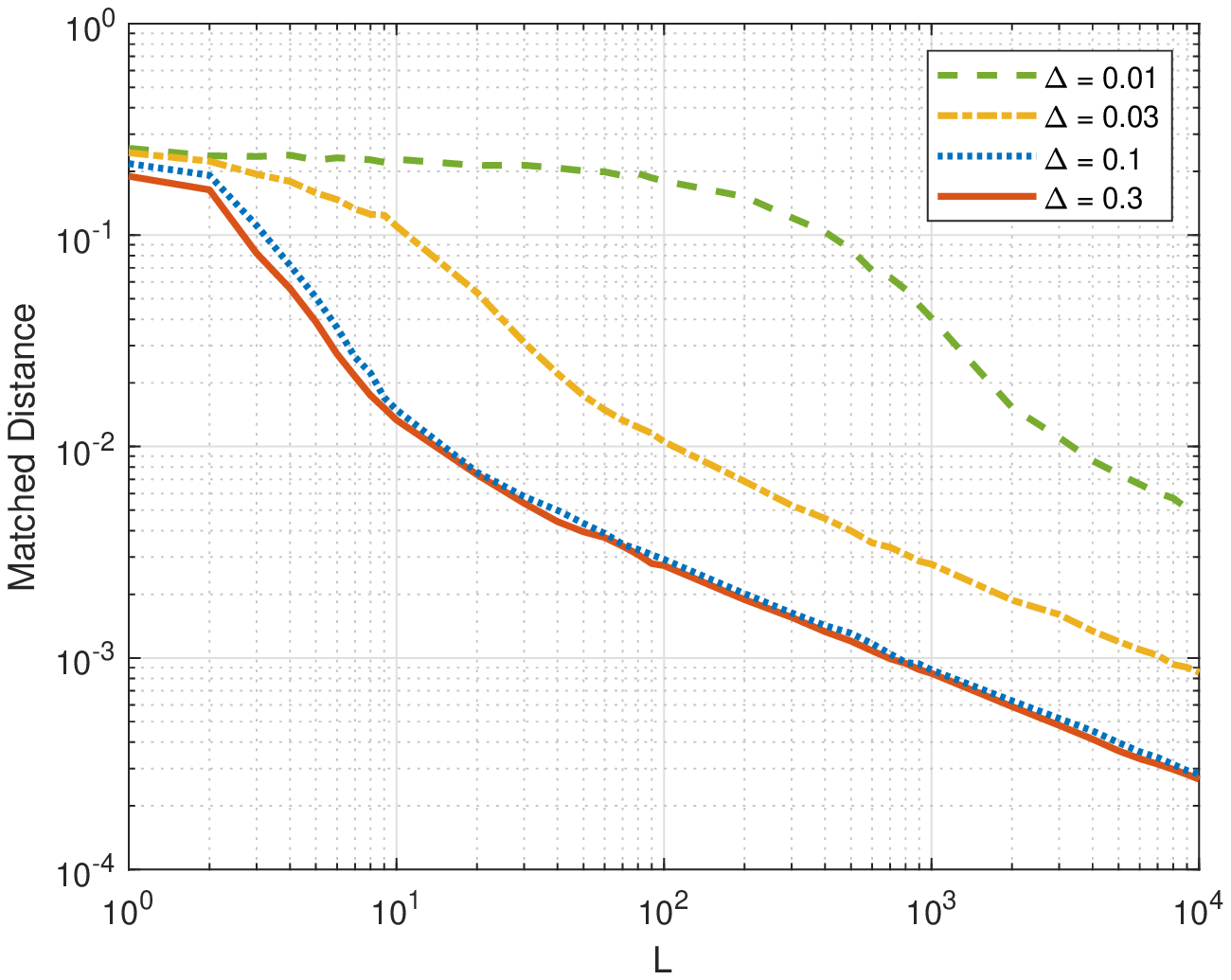}}
\subfigure[]{\label{fig:res_sigma}
\centering
\includegraphics[width=3.5in]{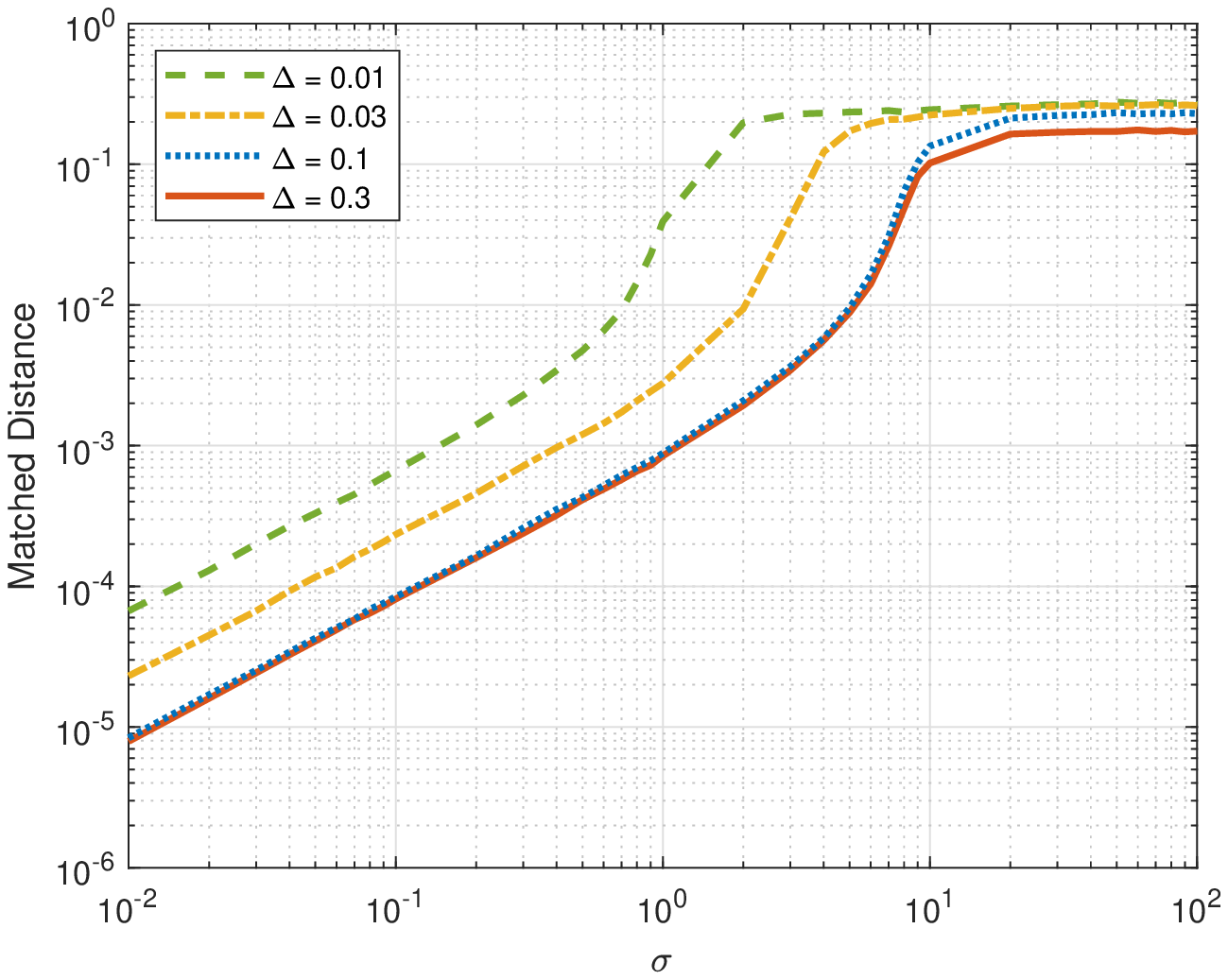}}
\caption{Results of matched distance of frequency estimation using ESPRIT with (a) fixed noise level $\sigma=1$ and varying number of snapshots $L$, and (b) fixed number of snapshots $L=1000$ and varying noise level $\sigma$.}
\end{figure}

In {\em Experiment 3}, we repeat {\em Experiment 1} and {\em Experiment 2} by varying the minimum frequency separation $\Delta$. The set of $K=3$ frequencies is given by $\lbra{0.1, 0.8 - \Delta, 0.8}$ where $\Delta\in \lbra{0.01, 0.03, 0.1,0.3}$. This means that the first and the third DOAs are fixed at $11.54^\circ$ and $-23.58^\circ$, respectively, while the second DOA varies from $-24.83^\circ$, $-27.39^\circ$, $-36.87^\circ$ to $-90^\circ$. Note that the case of $\Delta = 0.3$ recovers the setup in previous experiments and the case of $\Delta \leq \frac{1}{N}=0.1$ is usually referred to as the super-resolution regime. We first fix the noise level $\sigma=1$ and vary the number of snapshots $L$ and our simulation results are presented in Fig.~\ref{fig:res_L}. It is seen that a smaller frequency separation $\Delta$ results in a larger estimation error, especially in the super-resolution regime. The same scaling behavior is shown for all values of $\Delta$ given sufficiently large $L$, as predicted by Theorem \ref{thm:mdthm}. It is also seen that ESPRIT has higher resolution as $L$ increases, which is consistent with Corollary \ref{thm:resthm}. In Fig.~\ref{fig:res_sigma} we fix the number of snapshots $L=1000$ and vary the noise level $\sigma$. A similar behavior of ESPRIT is shown for all values of $\Delta$. It is also seen that a smaller noise level leads to a higher resolution of ESPRIT, which is consistent with our analysis.

\begin{figure}[htbp] \centerline{\includegraphics[width=3.5in]{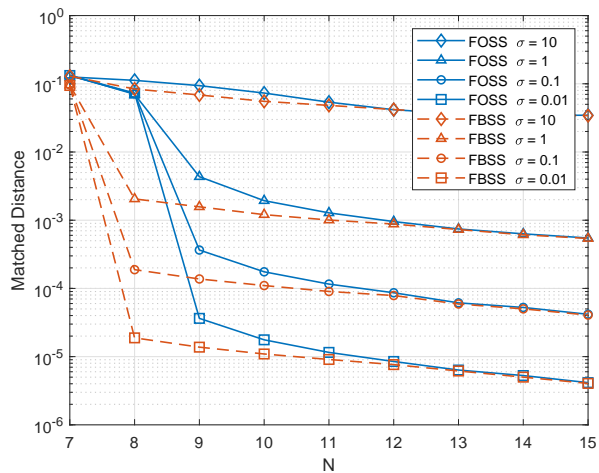}}
\caption{Results of matched distance of frequency estimation using FOSS-ESPRIT and FBSS-ESPRIT versus the per-snapshot sample size $N$.}
	\label{fig:ss_esprit}
\end{figure}

In {\em Experiment 4}, we consider the case of coherent sources, to be specific, $K=6$ sources of unit power are generated with a standard complex Gaussian distribution and are assigned to $G=3$ coherent groups with $g_1=3$, $g_2=2$ and $g_3=1$ (in each coherent group, the sources are identical up to global random phases). Their frequencies are given by the set $\cT=\lbra{0.1,0.2,0.5,0.6,0.7,0.9}$ that corresponds to the set of DOAs $\lbra{11,54^\circ, 23.58^\circ, -90^\circ, -53.13^\circ, -36.87^\circ, -11.54^\circ}$. We consider $L=1000$ snapshots and the noise level $\sigma\in\lbra{0.01,0.1,1,10}$. In this case, ESPRIT fails to localize the frequencies and so we turn to FOSS-ESPRIT and FBSS-ESPRIT. For both FOSS-ESPRIT and FBSS-ESPRIT, we fix $M=K+1=7$ and vary the smoothing parameter $P$ from $1$ to $9$ so that the per-snapshot sample size $N=M+P-1$ varies from $7$ to $15$. Note that FOSS-ESPRIT is exactly ESPRIT as $P=1$. Our numerical results are presented in Fig.~\ref{fig:ss_esprit}. It is seen that, as $\sigma\leq 1$, FOSS-ESPRIT stably estimates the frequencies as $N\geq 9$ (or $P\geq 3=g_1$) and FBSS-ESPRIT does so as $N\geq 8$ (or $P\geq 2$), which is consistent with our analysis. It is also seen that, as $\sigma\leq 1$, the matched distance shrinks by a magnitude as $\sigma$ does so for both FOSS-ESPRIT and FBSS-ESPRIT, implying that their frequency estimation errors scale with $\sigma$, as shown in Theorems \ref{thm:mdthm1} and \ref{thm:mdthm}. Finally, note that a significant performance gap exists between FOSS-ESPRIT and FBSS-ESPRIT when $N$ is small.

To sum up, it is shown by the numerical results that the scaling laws of the estimation errors of ESPRIT and SS-ESPRIT with respect to the snapshot number $L$ and the noise level $\sigma$ match well with our theoretical findings. We also note that the shown error bounds are conservative in their coefficients, which occurs partly due to the fact that the error bounds hold with overwhelming probability and partly due to limitations of current analysis techniques.

%
%
%
%
%
%
%
%

\section{Conclusion}
In this paper, we performed nonasymptotic analyses for ESPRIT and SS-ESPRIT. We showed that ESPRIT and SS-ESPRIT can stably estimate the frequencies with finite snapshots and finite SNR with overwhelming probability if and only if they localize the true frequencies with infinite snapshots or infinite SNR. For FBSS-ESPRIT, this occurs if and only if the frequencies can be uniquely identified from the data in the limiting case. Numerical results were provided that validate our theoretical findings.

MUSIC and ESPRIT are two prominent subspace methods, while we only considered ESPRIT and its variant SS-ESPRIT in the main context of this paper. In fact, similar conclusions can be drawn for MUSIC and SS-MUSIC by combining our error bounds on signal subspace estimation and previous analysis for single-snapshot MUSIC in \cite{liao2016music}. We provided the details in Appendix \ref{app:music}.

\appendix

\subsection{Why the Result in \cite[Theorem III.4]{li2021stability} is Incorrect} \label{sec:flaw}
In \cite{li2021stability}, heterogeneous noise is considered and in that case the data covariance matrix $\m{R}$ in \eqref{eq:R} becomes
\equ{\m{R} = \m{A}\m{\Sigma}\m{A}^H + \diag\lbra{\sigma_1^2,\dots,\sigma_N^2}, \label{eq:R2}}
where $\lbra{\sigma_j^2}$ are distinct noise powers. The main results of \cite{li2021stability} are derived based on its Theorem III.4 in which it is shown that the error of signal subspace estimation is bounded from above by a constant times $\frac{\max \sigma_j}{\sqrt{L}}$. If Theorem III.4 is true, then it immediately follows that the signal subspace estimation is consistent and recovers the range space of $\m{A}$ if $L\rightarrow \infty$. In other words, the range space of $\m{A}$ is exactly obtained from the eigen-decomposition of $\m{R}$, which evidently is not true in general for distinct $\lbra{\sigma_j^2}$. This error comes from \cite[Eq.~A.4]{li2021stability} that holds if $E_L$ and $\widehat{V}_1$ are independent, which however is not satisfied by noting that $\widehat{V}_1$ is obtained based on $E_L$. Note also that the above error remains to exist if i.i.d.~noise is assumed instead.

The above error has been fixed in \cite[version 2]{li2021stability} after the author of the present paper found the error and brought \cite{cai2018rate} to attention of the authors of \cite{li2021stability}, as acknowledged in \cite[version 2]{li2021stability}. Differently from our results for ESPRIT that provide upper bounds (with overwhelming probability) on the frequency estimation error, the results in \cite[version 2]{li2021stability}, derived based on \cite{cai2018rate}, turn to bound the expectation of the estimation error.

\subsection{Proof of Theorem \ref{thm:newdist}} \label{app:proof_newdist}
Let
{\lentwo\equa{\m{G}
&=& \m{M}\m{M}^H +\sigma^2 n\m{I} \notag\\
&=& \begin{bmatrix}\m{U} & \m{U}_{\perp} \end{bmatrix} \begin{bmatrix}\m{\Sigma}^2 & \\ & \m{\Sigma}_{\perp}^2 \end{bmatrix} \begin{bmatrix}\m{U}^H \\ \m{U}_{\perp}^H \end{bmatrix}  +\sigma^2 n\m{I},\\ \widehat{\m{G}}
&=& \widehat{\m{M}}\widehat{\m{M}}^H = \begin{bmatrix}\widehat{\m{U}} & \widehat{\m{U}}_{\perp} \end{bmatrix} \begin{bmatrix}\widehat{\m{\Sigma}}^2 & \\ & \widehat{\m{\Sigma}}_{\perp}^2 \end{bmatrix} \begin{bmatrix}\widehat{\m{U}}^H \\ \widehat{\m{U}}_{\perp}^H \end{bmatrix}.}
}Evidently, the eigenvectors of $\m{G}$ and $\widehat{\m{G}}$ are given by the left singular vectors of $\m{M}$ and $\widehat{\m{M}}$. The eigenvalues of $\m{G}$ satisfy that
\equ{\lambda_{r}\sbra{\m{G}} - \lambda_{r+1}\sbra{\m{G}} = \sigma_{r}^2 - \sigma_{r+1}^2.}
Moreover, the perturbation of $\widehat{\m{M}}$ from $\m{M}$ is given by
\equ{\widetilde{\m{E}}
= \widehat{\m{G}} - \m{G} = \m{M}\m{E}^H + \m{E}\m{M}^H + \m{E}\m{E}^H - \sigma^2n\m{I}. \label{eq:Etilde}}
Applying the Davis-Kahan $\sin\Theta$ theorem (Theorem \ref{thm:DK}), we have
\equ{\text{dist}\sbra{\widehat{\m{U}},\m{U}}\leq \frac{2\norm{\widetilde{\m{E}}}} {\lambda_r\sbra{\m{G}} - \lambda_{r+1}\sbra{\m{G}}}= \frac{2\norm{\widetilde{\m{E}}}} {\sigma_r^2 - \sigma_{r+1}^2} \label{eq:distinE2}}
if $\norm{\widetilde{\m{E}}}< 0.293\sbra{\sigma_r^2 - \sigma_{r+1}^2}$.

We next bound $\norm{\widetilde{\m{E}}}$ from above using random matrix theory. It follows from Lemma \ref{lem:Gausnorm} that
\equ{\begin{split}\norm{\m{E}\m{E}^H - \sigma^2n\m{I}}
&= \sigma^2n\norm{\frac{1}{\sigma^2n}\m{E}\m{E}^H - \m{I}} \\
&\leq \sigma^2n\mbra{2\cdot 2\sqrt{\frac{p}{n}} + \sbra{2\sqrt{\frac{p}{n}}}^2}\\
&= 4\sigma^2\sbra{\sqrt{pn} + p}\\
&\leq 8\sigma^2\max\lbra{\sqrt{pn}, p} \end{split} \label{eq:EEmInorm}}
with probability at least $1-2e^{-\frac{p}{2}}$, where $u=\sqrt{\frac{p}{n}}$ is used. To bound $\norm{\m{M}\m{E}^H}$, note that the columns of the $p\times p$ matrix $\m{M}\m{E}^H$ are i.i.d.~Gaussian with zero mean and covariance $\sigma^2\m{M}\m{M}^H$. It follows that $\m{M}\m{E}^H$ can be written as $\sigma\sbra{\m{M}\m{M}^H}^{\frac{1}{2}}\m{E}'^H$, where $\m{E}'$ is $p\times p$ and i.i.d.~standard Gaussian. Consequently,
\equ{\begin{split}\norm{\m{M}\m{E}^H}
&= \norm{\sigma\sbra{\m{M}\m{M}^H}^{\frac{1}{2}}\m{E}'^H} \\
&\leq \sigma\norm{\sbra{\m{M}\m{M}^H}^{\frac{1}{2}}} \norm{\m{E}'}\\
&= \sigma\sigma_1 \norm{\m{E}'}\\
&\leq 3\sigma\sigma_1\sqrt{p} \end{split} \label{eq:MEnorm}}
with probability at least $1-e^{-\frac{p}{2}}$, where the last inequality follows from Lemma \ref{lem:Gausnorm} by setting $u=\sqrt{p}$. Combining \eqref{eq:Etilde}, \eqref{eq:EEmInorm} and \eqref{eq:MEnorm}, we have
\equ{\begin{split}\norm{\widetilde{\m{E}}}
&\leq 2\norm{\m{M}\m{E}^H} + \norm{\m{E}\m{E}^H - \sigma^2n\m{I}} \\
&\leq 6\sigma\sigma_1\sqrt{p} + 8\sigma^2\max\lbra{\sqrt{pn}, p} \end{split} \label{eq:Etildenorm}}
with probability at least $1-3e^{-\frac{p}{2}}$. Substituting \eqref{eq:Etildenorm} into \eqref{eq:distinE2}, we obtain \eqref{eq:newdist}, completing the proof.

\subsection{Proof of Theorem \ref{thm:dist_noSS}} \label{app:proof_noSS}
To apply Theorem \ref{thm:newdist}, we identify that
\lentwo{\equa{\m{M}
&=& \m{A}\m{S},\\ \widehat{\m{M}}
&=& \m{Y}= \m{A}\m{S} + \m{E}}
}with $p=N$, $n=L$ and $r=K$. It follows from Theorem \ref{thm:newdist} that if the upper bound below is less than $0.586$, then with probability at least $1-3e^{-\frac{N}{2}}$ we have
\equ{\begin{split}
&\text{dist}\sbra{\widehat{\m{U}},\m{U}}\\
&\leq \frac{12\sigma\sigma_1\sbra{\m{A}\m{S}}\sqrt{N} + 16\sigma^2\max\lbra{\sqrt{NL}, N}} {\sigma_K^2\sbra{\m{A}\m{S}} - \sigma_{K+1}^2\sbra{\m{A}\m{S}}}\\
&\leq \frac{12\sigma\norm{\m{A}}\norm{\m{S}}\sqrt{N} + 16\sigma^2\max\lbra{\sqrt{NL}, N}} {\sigma_K^2\sbra{\m{A}}\sigma_K^2\sbra{\m{S}}}\\
&= \frac{12\sigma\norm{\m{A}}{\norm{L\m{\widehat{\Sigma}}}}^{\frac{1}{2}}\sqrt{N} + 16\sigma^2\max\lbra{\sqrt{NL}, N}} {\sigma_K^2\sbra{\m{A}} \lambda_K\sbra{L\widehat{\m{\Sigma}}}}\\
&= \frac{12\sigma\norm{\m{A}}{\norm{\m{\widehat{\Sigma}}}}^{\frac{1}{2}}\sqrt{\frac{N}{L}} + 16\sigma^2\max\lbra{\sqrt{\frac{N}{L}}, \frac{N}{L}}} {\sigma_K^2\sbra{\m{A}} \lambda_K\sbra{\widehat{\m{\Sigma}}}}, \end{split}\label{eq:dist_noSS2}}
completing the proof. Note that in \eqref{eq:dist_noSS2} we used the fact that $\sigma_1\sbra{\m{A}\m{S}} \leq \norm{\m{A}}\norm{\m{S}}$, $\sigma_K\sbra{\m{A}\m{S}} \geq \sigma_K\sbra{\m{A}}\sigma_K\sbra{\m{S}}$ and $\widehat{\m{\Sigma}} = \frac{1}{L}\m{S}\m{S}^H$.

\subsection{Proof of Theorem \ref{thm:dist_SS}} \label{app:proof_SS}
It follows from \eqref{eq:RSShat} and \eqref{eq:YSS} that the smoothed sample covariance matrix is given by
\equ{\begin{split}\widehat{\m{R}}_{\text{SS}}
&= \frac{1}{P}\sum_{p=1}^P \m{A}_M\m{Z}^{p-1}\widehat{\m{\Sigma}}\m{Z}^{1-p}\m{A}_M^H \\
&\quad+ \frac{1}{PL}\sum_{p=1}^P\m{A}_M\m{Z}^{p-1}\m{S} \m{E}_{(p)}^H \\
&\quad + \frac{1}{PL}\sum_{p=1}^P\m{E}_{(p)}\m{S}^H\m{Z}^{1-p}\m{A}_M^H + \frac{1}{PL}\sum_{p=1}^P\m{E}_{(p)}\m{E}_{(p)}^H\\
&= \m{A}_M\widehat{\m{\Sigma}}_{\text{SS}}\m{A}_M^H + \frac{1}{PL}\sum_{p=1}^P\m{A}_M\m{Z}^{p-1}\m{S} \m{E}_{(p)}^H \\
&\quad + \frac{1}{PL}\sum_{p=1}^P\m{E}_{(p)}\m{S}^H\m{Z}^{1-p}\m{A}_M^H + \frac{1}{PL}\sum_{p=1}^P\m{E}_{(p)}\m{E}_{(p)}^H, \end{split}}
where the last equality follows from derivations similar to those in \eqref{eq:sSigma} and \eqref{eq:sSigma2}. Note that $\m{U}$ is the eigen-subspace of
\equ{\begin{split}\widetilde{\m{R}}_{\text{SS}}
&= \m{A}_M\widehat{\m{\Sigma}}_{\text{SS}}\m{A}_M^H + \sigma^2\m{I} \end{split}}
if $\widehat{\m{\Sigma}}_{\text{SS}}$ is positive definite. We next carry out perturbation analysis between $\widehat{\m{R}}_{\text{SS}}$ and $\widetilde{\m{R}}_{\text{SS}}$. The gap between the $K$th and $(K+1)$st eigenvalues of $\widetilde{\m{R}}_{\text{SS}}$ equals
\equ{\begin{split}\lambda_{K}\sbra{\widetilde{\m{R}}_{\text{SS}}} - \lambda_{K+1}\sbra{\widetilde{\m{R}}_{\text{SS}}}
&= \lambda_K\sbra{\m{A}_M\widehat{\m{\Sigma}}_{\text{SS}}\m{A}_M^H} \\
&\geq \sigma_K^2\sbra{\m{A}_M}\lambda_K\sbra{\widehat{\m{\Sigma}}_{\text{SS}}}. \end{split}}
Moreover, the perturbation of $\widehat{\m{R}}_{\text{SS}}$ from $\widetilde{\m{R}}_{\text{SS}}$ is given by
\equ{\begin{split}\widetilde{\m{E}}
&= \widehat{\m{R}}_{\text{SS}} - \widetilde{\m{R}}_{\text{SS}} \\
&= \frac{1}{PL}\sum_{p=1}^P\sbra{\m{A}_M\m{Z}^{p-1}\m{S} \m{E}_{(p)}^H + \m{E}_{(p)}\m{S}^H\m{Z}^{1-p}\m{A}_M^H} \\
&\quad + \frac{1}{PL}\sum_{p=1}^P\m{E}_{(p)}\m{E}_{(p)}^H - \sigma^2\m{I}. \end{split} \label{eq:Etilde2}}
Applying the Davis-Kahan $\sin\Theta$ theorem (Theorem \ref{thm:DK}), we have
\equ{\begin{split}\text{dist}\sbra{\widehat{\m{U}},\m{U}}
&\leq \frac{2\norm{\widetilde{\m{E}}}} {\lambda_{K}\sbra{\widetilde{\m{R}}_{\text{SS}}} - \lambda_{K+1}\sbra{\widetilde{\m{R}}_{\text{SS}}} }\\
&\leq \frac{2\norm{\widetilde{\m{E}}}} {\sigma_K^2\sbra{\m{A}_M}\lambda_K\sbra{\widehat{\m{\Sigma}}_{\text{SS}}}} \end{split}\label{eq:distinE}}
if the upper bound above is no greater than $0.586$.

We next bound $\norm{\widetilde{\m{E}}}$ from above. Clearly, it follows from \eqref{eq:Etilde2} that
\equ{\begin{split}\norm{\widetilde{\m{E}}}
&\leq \frac{2}{L}\max_p\norm{ \m{A}_M\m{Z}^{p-1}\m{S} \m{E}_{(p)}^H}\\
&\quad +\max_p\norm{\frac{1}{L}\m{E}_{(p)}\m{E}_{(p)}^H - \sigma^2 \m{I}}\\
&\leq \frac{2}{L}\max_p\norm{\m{A}_M}\norm{\m{S} \m{E}_{(p)}^H} \\
&\quad + \sigma^2\max_p\norm{\frac{1}{L\sigma^2}\m{E}_{(p)}\m{E}_{(p)}^H - \m{I}}. \end{split} }
Note that the columns of the $K\times M$ matrix $\m{S} \m{E}_{(p)}^H$ are i.i.d.~multivariate Gaussian with zero mean and variance $\sigma^2\m{S}\m{S}^H=L\sigma^2\widehat{\m{\Sigma}}$. It follows from derivations similar to those in \eqref{eq:MEnorm} that for each $p$,
\equ{\norm{\m{S} \m{E}_{(p)}^H} \leq 3\sigma\sqrt{LM\norm{\widehat{\m{\Sigma}}}}}
with probability at least $1-e^{-\frac{M}{2}}$. Moreover, similarly to \eqref{eq:EEmInorm}, we have that for each $p$,
\equ{\norm{\frac{1}{L\sigma^2}\m{E}_{(p)}\m{E}_{(p)}^H - \m{I}} \leq 8\max\lbra{\sqrt{\frac{M}{L}},\frac{M}{L}}}
with probability at least $1-2e^{-\frac{M}{2}}$. As a result,
\equ{\norm{\widetilde{\m{E}}} \leq 6\sigma\norm{\m{A}_M}\norm{\widehat{\m{\Sigma}}}^{\frac{1}{2}}\sqrt{\frac{M}{L}} + 8\sigma^2 \max\lbra{\sqrt{\frac{M}{L}}, \frac{M}{L}} \label{eq:boundtE}}
with probability at least $1-3Pe^{-\frac{M}{2}}$.

Inserting \eqref{eq:boundtE} into \eqref{eq:distinE}, we obtain \eqref{eq:dist_noSS}, completing the proof.

\subsection{Proof of Theorem \ref{thm:dist_FBSS}} \label{app:proof_FBSS}
Recall \eqref{eq:RSShat}. Then, we have
\equ{\begin{split}\widehat{\m{R}}'_{\text{SS}}
&= \frac{1}{2}\sbra{\widehat{\m{R}}_{\text{SS}} + \m{J}\overline{\widehat{\m{R}}_{\text{SS}}}\m{J}} =\frac{1}{LP} \m{Y}'_{\text{SS}}{\m{Y}}_{\text{SS}}'^{H}, \end{split} \label{eq:RSShatp}}
and thus $\widehat{\m{U}}$ is the left singular subspace of
\equ{\begin{split}\m{Y}'_{\text{SS}}
&= \frac{1}{\sqrt{2}}\mbra{\m{Y}_{(1)},\; \dots,\;\m{Y}_{(P)}, \m{J}\overline{\m{Y}_{(1)}},\; \dots,\;\m{J}\overline{\m{Y}_{(P)}}}\\
&= [\m{A}\m{S}]_{\text{SS}}' + \m{E}_{\text{SS}}', \end{split} \label{eq:YSSp}}
where $[\m{A}\m{S}]_{\text{SS}}',\m{E}_{\text{SS}}'$ are defined similarly as $\m{Y}'_{\text{SS}} $. Note also that $\m{U}$ is the left singular subspace of
\equ{\begin{split}
&[\m{A}\m{S}]_{\text{SS}}'\\
&= \frac{1}{\sqrt{2}}\left[\m{A}_M\m{S},\;\dots,\; \m{A}_M\m{Z}^{P-1}\m{S}, \; \m{J}\overline{\m{A}_M\m{S}},\;\dots,\right.\\
&\phantom{\m{A}_M\m{S},\;\dots,\; \m{A}_M\m{Z}^{P-1}\m{S}, \; \m{J}\overline{\m{A}_M\m{S}},\;\dots } \left. \m{J}\overline{\m{A}_M\m{Z}^{P-1}\m{S}}\right]\\
&= \frac{1}{\sqrt{2}}\m{A}_M\mbra{\m{S},\;\dots,\; \m{Z}^{P-1}\m{S}, \m{Z}^{1-M}\overline{\m{S}},\;\dots,\; \m{Z}^{2-P-M}\overline{\m{S}}} \end{split}}
if $\mbra{\m{A}\m{S}}_{\text{SS}}'$ has rank $K$. The proof is completed by repeating the proof of Theorem \ref{thm:dist_SS} based on the perturbation model in \eqref{eq:YSSp}. We will omit the details.

\subsection{Proof of Theorem \ref{thm:mdthm}} \label{app:proof_mdthm}
Let $\m{G}$ be a matrix of full column rank and satisfy that $\m{\Sigma} = \m{G}\m{G}^H$. It follows that
\equ{\m{S} = \m{G}\check{\m{S}},}
where $\check{\m{S}}$ is $\rank\sbra{\m{\Sigma}}\times L$ and i.i.d.~standard Gaussian.
Since $L\geq 16 \rank\sbra{\m{\Sigma}}$, we take $u=\frac{1}{4}\sqrt{L}$ and it follows from Lemma \ref{lem:Gausnorm} that
\lentwo{\equa{\lambda_1\sbra{\check{\m{S}}\check{\m{S}}^H}
&\leq& \sbra{\sqrt{L}+\sqrt{\rank\sbra{\m{\Sigma}}} + \frac{1}{4}\sqrt{L} }^2 \leq \frac{9}{4}L, \label{eq:lambda1upp}\\ \lambda_K\sbra{\check{\m{S}}\check{\m{S}}^H}
&\geq& \sbra{\sqrt{L}-\sqrt{\rank\sbra{\m{\Sigma}}} - \frac{1}{4}\sqrt{L} }^2 \geq \frac{1}{4}L, \label{eq:lambdaKlow}}
}each with probability at least $1-e^{-\frac{L}{32}}$. Conditioning on \eqref{eq:lambda1upp} and \eqref{eq:lambdaKlow}, consequently, we have
\lentwo{\equa{\frac{1}{4} \m{\Sigma} \leq \widehat{\m{\Sigma}}
&=& \m{G}\sbra{\frac{1}{L}\check{\m{S}}\check{\m{S}}^H}\m{G}^H
\leq \frac{9}{4} \m{\Sigma}, \label{eq:Sigmatint}\\ \norm{\widehat{\m{\Sigma}}}
&\leq& \frac{9}{4} \norm{\m{\Sigma}},\label{eq:Sigmanupp} \\ \lambda_K\sbra{\widehat{\m{\Sigma}}}
&\geq& \frac{1}{4}\lambda_K\sbra{\m{\Sigma}}. \label{eq:hplow}}
}Substituting \eqref{eq:Sigmanupp} and \eqref{eq:hplow} into \eqref{eq:dist_noSS} and using the assumption $L\geq \max\lbra{N, 16 K}$, we have
\equ{\begin{split}\text{dist}\sbra{\widehat{\m{U}},\m{U}}
&\leq \frac{18\sigma\norm{\m{A}}\norm{\m{\Sigma}}^{\frac{1}{2}}\sqrt{\frac{N}{L}} + 16\sigma^2 \sqrt{\frac{N}{L}}} {\frac{1}{4}\sigma_K^2\sbra{\m{A}} \lambda_K\sbra{\m{\Sigma}}}\\
&\leq 72\sqrt{\frac{N}{L}}\cdot \frac{\sigma\norm{\m{A}}\norm{\m{\Sigma}}^{\frac{1}{2}} + \sigma^2 } {\sigma_K^2\sbra{\m{A}} \lambda_K\sbra{\m{\Sigma}}} \end{split}\label{eq:distinE3}}
with probability at least $1-3e^{-\frac{N}{2}}-2e^{-\frac{L}{32}}$. Inserting \eqref{eq:distinE3} into \eqref{eq:mdlem} yields \eqref{eq:mdthm_noSS}.

To derive \eqref{eq:mdthm_SS} and \eqref{eq:mdthm_FBSS}, we make use of \eqref{eq:Sigmatint} and the Schur product theorem (see Theorem \ref{thm:schur}) and obtain that
\lentwo{\equa{\widehat{\m{\Sigma}}_{\text{SS}}
&=& \widehat{\m{\Sigma}} \odot \m{C}_P
\geq \frac{1}{4}\m{\Sigma} \odot \m{C}_P = \frac{1}{4}\m{\Sigma}_{\text{SS}},\\ \widehat{\m{\Sigma}}'_{\text{SS}}
&=& \frac{1}{2} \sbra{\widehat{\m{\Sigma}}_{\text{SS}} + \m{Z}^{1-M} \overline{\widehat{\m{\Sigma}}_{\text{SS}}}\m{Z}^{M-1} } \notag\\
&\geq& \frac{1}{2} \sbra{\frac{1}{4}\m{\Sigma}_{\text{SS}} + \frac{1}{4}\m{Z}^{1-M} \overline{\m{\Sigma}_{\text{SS}}}\m{Z}^{M-1} } \\
&=& \frac{1}{4}\m{\Sigma}'_{\text{SS}},\notag
}}yielding
\lentwo{\equa{\lambda_K\sbra{\widehat{\m{\Sigma}}_{\text{SS}}}
&\geq&  \frac{1}{4}\lambda_K\sbra{\m{\Sigma}_{\text{SS}}}, \label{eq:SigmatSSupp}\\ \lambda_K\sbra{\widehat{\m{\Sigma}}'_{\text{SS}}}
&\geq& \frac{1}{4}\lambda_K\sbra{\m{\Sigma}'_{\text{SS}}}. \label{eq:SigmattSSupp}
}}Using \eqref{eq:SigmatSSupp}, \eqref{eq:SigmattSSupp}, \eqref{eq:dist_SS} and \eqref{eq:dist_FBSS}, instead of \eqref{eq:hplow} and \eqref{eq:mdlem}, and repeating our previous arguments conclude the proof. We will omit the details.

\subsection{Proof of Proposition \ref{prop:newHPlb}} \label{app:proof_newHPlb}
For simplicity we denote $\m{C}=\m{C}_P$ and partition $\m{C}$ into a $G\times G$ block matrix, $\m{C} = [\m{C}_{ij}]_{G\times G}$, as $\m{\Sigma}$ in \eqref{eq:Sigmadec}. It suffices to show that
\begin{align}
\lambda_{\text{min}}\sbra{\m{\Sigma}\odot \m{C}}
&\geq \lambda_{\text{min}}\sbra{\check{\m{\Sigma}}} \min_j \lbra{\lambda_{\text{min}}\sbra{\m{C}_{{jj}}} \min_l \abs{v_{jl}}^2}, \label{eq:newlb2} \\ \lambda_{\text{min}}\sbra{\check{\m{\Sigma}}}
&= \lambda_{G}\sbra{\m{\Sigma}}. \label{eq:lambdaGSig}
\end{align}

To show \eqref{eq:newlb2}, observe that
\equ{\begin{split}
&\m{\Sigma} - \lambda_{\text{min}}\sbra{\check{\m{\Sigma}}} \diag\sbra{\m{v}_1\m{v}_1^H,\dots,\m{v}_G\m{v}_G^H} \\
&= \diag\sbra{\m{v}_1,\dots,\m{v}_G} \sbra{\check{\m{\Sigma}} - \lambda_{\text{min}}\sbra{\check{\m{\Sigma}}}\m{I}} \diag^H\sbra{\m{v}_1,\dots,\m{v}_G} \end{split}}
is positive semidefinite, and so is
\equ{\begin{split}
&\mbra{\m{\Sigma} - \lambda_{\text{min}}\sbra{\check{\m{\Sigma}}} \diag\sbra{\m{v}_1\m{v}_1^H,\dots,\m{v}_G\m{v}_G^H}}\odot\m{C} \\
&= \m{\Sigma}\odot\m{C} - \lambda_{\text{min}}\sbra{\check{\m{\Sigma}}} \diag\sbra{\m{v}_1\m{v}_1^H,\dots,\m{v}_G\m{v}_G^H}\odot\m{C} \end{split}}
by applying the Schur product theorem (Theorem \ref{thm:schur}), where $\diag\sbra{\m{v}_1\m{v}_1^H,\dots,\m{v}_G\m{v}_G^H}$ is a block diagonal matrix. Consequently,
\equ{\begin{split}
&\lambda_{\text{min}}\sbra{\m{\Sigma}\odot\m{C}}\\
&\geq \lambda_{\text{min}}\sbra{\lambda_{\text{min}}\sbra{\check{\m{\Sigma}}} \diag\sbra{\m{v}_1\m{v}_1^H,\dots, \m{v}_G\m{v}_G^H}\odot\m{C}}\\
&= \lambda_{\text{min}}\sbra{\check{\m{\Sigma}}} \lambda_{\text{min}}\sbra{ \diag\sbra{\m{v}_1\m{v}_1^H\odot\m{C}_{11},\dots, \m{v}_G\m{v}_G^H \odot\m{C}_{GG}}}\\
&= \lambda_{\text{min}}\sbra{\check{\m{\Sigma}}} \min_j \lambda_{\text{min}} \sbra{\m{v}_j\m{v}_j^H\odot\m{C}_{jj}}\\
&\geq \lambda_{\text{min}}\sbra{\check{\m{\Sigma}}} \min_j \lbra{\lambda_{\text{min}} \sbra{\m{C}_{jj}} \cdot \min_l \abs{v_{jl}}^2},
\end{split}}
where again the last inequality follows from the Schur product theorem.

We next show \eqref{eq:lambdaGSig} to complete the proof. Suppose that $\lambda, \m{u}$ form an eigenvalue-eigenvector pair of $\check{\m{\Sigma}}$, i.e., $\check{\m{\Sigma}}\m{u} = \lambda\m{u}$. It immediately follows that
\equ{\begin{split}\m{\Sigma} \sbra{\diag\sbra{\m{v}_1,\dots,\m{v}_G}\m{u}}
&= \diag\sbra{\m{v}_1,\dots,\m{v}_G} \check{\m{\Sigma}} \m{u}\\
&= \lambda \diag\sbra{\m{v}_1,\dots,\m{v}_G} \m{u}, \end{split}}
and thus $\lambda, \diag\sbra{\m{v}_1,\dots,\m{v}_G}\m{u}$ form an eigenvalue-eigenvector pair of $\m{\Sigma}$. Therefore, the eigenvalues of $\m{\Sigma}$ are either those of $\check{\m{\Sigma}}$ or zero, resulting in \eqref{eq:lambdaGSig}.

\subsection{Extensions to MUSIC and SS-MUSIC} \label{app:music}
We first introduce the MUSIC algorithm. Differently from ESPRIT, MUSIC uses the noise subspace $\m{U}_{\perp}$, where $\m{U}$ is the signal subspace. Let $\cP_{\m{U}_{\perp}}$ be the orthogonal projection onto the noise subspace and define the noise subspace correlation function
\equ{\cR(f) = \frac{\norm{\cP_{\m{U}_{\perp}}\m{a}\sbra{f}}}{\norm{\m{a}\sbra{f}}},}
where $\m{a}\sbra{f} = \mbra{1,e^{i2\pi f},\dots,e^{i2\pi (N-1)f}}^T$.
Since $\m{U}$ and $\m{A}$ share the same range space, we have that $\cR(f)$ vanishes if and only if $f$ takes value in the frequency set $\cT$. In practice, $\m{U},\m{U}_{\perp}$ are replaced by their estimates $\widehat{\m{U}},\widehat{\m{U}}_{\perp}$, resulting in the estimated correlation function $\widehat{\cR}(f)$. The frequencies are estimated from the smallest $K$ local minima of $\widehat{\cR}(f)$ or equivalently, the highest $K$ peaks of the MUSIC spectral function given by $\frac{1}{\mbra{\widehat{\cR}(f)}^2}$.

We next revisit the analysis of single-snapshot MUSIC in \cite{liao2016music}. It is shown there that the frequency estimation error is determined by the error
\equ{\max_{f\in[0,1)}\abs{\widehat{\cR}(f) - \cR(f)}. \label{eq:fonT}}
If $\cR(f)$ is stably estimated, then so are the frequencies under mild conditions. Readers are referred to \cite{liao2016music} for details. Consequently, we only consider the stability of the error in \eqref{eq:fonT} with respect to noise hereafter. To do so, it is shown in \cite[Theorem 3]{liao2016music} that for $f\in[0,1)$,
\equ{\begin{split}\abs{\widehat{\cR}(f) - \cR(f)}
&= \frac{\abs{\norm{\cP_{\widehat{\m{U}}_{\perp}}\m{a}\sbra{f}}- \norm{\cP_{\m{U}_{\perp}}\m{a}\sbra{f}} }}{\sqrt{N}} \\
&\leq \frac{\norm{\sbra{\cP_{\m{U}_{\perp}} - \cP_{\widehat{\m{U}}_{\perp}}}\m{a}\sbra{f}}} {\sqrt{N}} \\
&\leq \norm{\cP_{\m{U}_{\perp}} - \cP_{\widehat{\m{U}}_{\perp}}} \end{split} \label{eq:fonT2}}
and the distance $\norm{\cP_{\m{U}_{\perp}} - \cP_{\widehat{\m{U}}_{\perp}}} $ is bounded from above by a constant times the noise level $\sigma$, concluding the stability of the correlation function $\widehat{\cR}\sbra{f}$ and frequency estimation of single-snapshot MUSIC.

Now we are ready to provide our analysis of the multiple-snapshot MUSIC and SS-MUSIC algorithms. Like ESPRIT, the only difference between the single-snapshot and multiple-snapshot MUSIC is the way of estimating the signal or noise subspace. Consequently, the stability of frequency estimation in our case can also be concluded by the stability of the correlation function $\widehat{\cR}\sbra{f}$, as in \cite{liao2016music}. To this end, we measure the error of the correlation function in \eqref{eq:fonT} by the error of signal subspace estimation $\text{dist}\sbra{\widehat{\m{U}},\m{U}}$ with
\equ{\max_{f\in[0,1)}\abs{\widehat{\cR}(f) - \cR(f)} \leq \text{dist}\sbra{\widehat{\m{U}}, \m{U}}, \label{eq:fonT3}}
by combining \eqref{eq:fonT2} and the identity
\equ{\norm{\cP_{\m{U}_{\perp}} - \cP_{\widehat{\m{U}}_{\perp}}} = \norm{\cP_{\m{U}} - \cP_{\widehat{\m{U}}}} = \text{dist}\sbra{\widehat{\m{U}}, \m{U}},}
where the first equality holds since $\cP_{\m{U}_{\perp}}+\cP_{\m{U}}$ is identity and the second follows from \eqref{eq:distdef2}. The stability of $\widehat{\cR}(f)$ is concluded by combining \eqref{eq:fonT3} and our Theorem \ref{thm:dist_noSS}. Similar conclusions can be drawn for FOSS-MUSIC and FBSS-MUSIC by combining \eqref{eq:fonT3} and Theorems \ref{thm:dist_SS} and \ref{thm:dist_FBSS} respectively.

\section*{Acknowledgment}
The author would like to thank Mr.~Kaijie Wang for helping preparing the figures, Prof.~Peter Stoica for helpful comments on an earlier draft of this paper, and the anonymous reviewers for their valuable comments that improved the quality of the paper.



\end{document}